\documentclass[twocolumn]{aastex63}

\usepackage{natbib}
\usepackage[utf8]{inputenc}

\hypersetup{linkcolor=red,citecolor=cyan,filecolor=cyan}

\newcommand{\msun}{\mbox{$\rm M_\odot$}}

\newcommand{\metallicity}{\mbox{$\rm log(\it Z/Z_{\odot}\rm )$}}

\newcommand{\nh}{\mbox{{\sc \small NewHorizon}}}
\newcommand{{\gal}}{\mbox{{\sc \small Galactica}}}

\submitjournal{ApJ}
\shortauthors{Park et al.}
\graphicspath{{./}{figures/}}

\usepackage{lipsum}
\usepackage{amsmath}
\usepackage{comment}
\usepackage{multirow}
\usepackage{enumitem}
\usepackage{threeparttable}
\usepackage{hyperref}
\usepackage{changepage}
\usepackage{amsmath}

\begin{document}
\inputencoding{utf8}
\title{Exploring the origin of thick disks using the NewHorizon and Galactica simulations}

\author[0000-0002-8435-9402]{Minjung J. Park}
\altaffiliation{Present address: Center for Astrophysics $|$ Harvard \& Smithsonian. minjung.park@cfa.harvard.edu}
\affiliation{Department of Astronomy and Yonsei University Observatory, Yonsei University, Seoul 03722, Republic of Korea}

\author[0000-0002-4556-2619]{Sukyoung K. Yi}
\altaffiliation{Email address: yi@yonsei.ac.kr}
\affil{Department of Astronomy and Yonsei University Observatory, Yonsei University, Seoul 03722, Republic of Korea}

\author{Sebastien Peirani}
\affil{Observatoire de la Côte d’Azur, CNRS, Laboratoire Lagrange, Bd de l’Observatoire,Université Côte d’Azur, CS 34229, 06304 Nice Cedex 4, France}
\affiliation{Institut d’Astrophysique de Paris, Sorbonne Université, CNRS, UMR 7095, 98 bis bd Arago, 75014 Paris, France}

\author[0000-0003-0695-6735]{Christophe Pichon}
\affiliation{Institut d’Astrophysique de Paris, Sorbonne Université, CNRS, UMR 7095, 98 bis bd Arago, 75014 Paris, France}
\affiliation{Korea Institute for Advanced Study, 85 Hoegiro, Dongdaemun-gu, 02455 Seoul, Republic of Korea}

\author[0000-0003-0225-6387]{Yohan Dubois}
\affil{Institut d’Astrophysique de Paris, Sorbonne Université, CNRS, UMR 7095, 98 bis bd Arago, 75014 Paris, France}

\author[0000-0001-7229-0033]{Hoseung Choi}
\affiliation{Department of Astronomy and Yonsei University Observatory, Yonsei University, Seoul 03722, Republic of Korea}

\author{Julien Devriendt}
\affil{Dept of Physics, University of Oxford, Keble Road, Oxford OX1 3RH, UK}

\author{Sugata Kaviraj}
\affil{Centre for Astrophysics Research, School of Physics, Astronomy and Mathematics, University of Hertfordshire, College Lane, Hatfield AL10 9AB, UK}

\author[0000-0002-3950-3997]{Taysun Kimm}
\affil{Department of Astronomy and Yonsei University Observatory, Yonsei University, Seoul 03722, Republic of Korea}

\author{Katarina Kraljic}
\affil{Institute for Astronomy, University of Edinburgh, Royal Observatory, Blackford Hill, Edinburgh EH9 3HJ, UK}
\affil{Aix Marseille Université, CNRS, CNES, UMR 7326, Laboratoire d’Astrophysique de Marseille, Marseille, France}

\author[0000-0002-3216-1322]{Marta Volonteri}
\affil{Institut d’Astrophysique de Paris, Sorbonne Université, CNRS, UMR 7095, 98 bis bd Arago, 75014 Paris, France}

\inputencoding{utf8}
\begin{abstract}

Ever since a thick disk was proposed to explain the vertical distribution of the Milky Way disk stars, its origin has been a recurrent question. We aim to answer this question by inspecting 19 disk galaxies with stellar mass greater than $10^{10}$ \msun\ in recent cosmological high-resolution zoom-in simulations: {\gal} and {\nh}. The thin and thick disks are reasonably reproduced by the simulations with scale heights and luminosity ratios as  observed. 
We then spatially classify the thin and thick disks and find that the thick disk stars are older, metal-poorer, kinematically-hotter, and higher in accreted star fraction, while both disks are dominated by the stars formed in situ. Half of the in-situ stars in the thick disks are formed before the galaxies develop their disks, and the rest are formed in spatially and kinematically thinner disks and then thickened with time by heating. 
However, the 19 galaxies have various properties and evolutionary routes, highlighting the need for statistically-large samples to draw general conclusions.
We conclude from our simulations that the thin and thick disk components are not entirely distinct in terms of formation processes, but rather markers of the evolution of galactic disks. 
Moreover, as the combined result of the thickening of the existing disk stars and the continued formation of young thin-disk stars, the vertical distribution of stars does not change much after the disks settle, pointing to the modulation of both orbital diffusion and star formation by the same confounding factor: the proximity of galaxies to marginal stability. 
\end{abstract}



\section{Introduction} 
\label{sec:intro}

The idea of the two-component disk structure for the Milky Way (MW) galaxy was first proposed by \cite{Gilmore1983GalacticDisc} to explain the vertical distribution of disk stars. 
They measured the number density of resolved stars in the solar neighborhood and found that the vertical structure of the MW disk is well described by a double exponential profile: the thin disk with a scale height of $\sim\rm 300\,pc$ and the thick disk with a scale height of $\sim\rm 1450\,pc$ \citep[see also][]{Juric2008TheDistribution, Ivezic2008TheMetallicity}. 
The level of contribution from thick disk stars to the solar neighborhood is suggested to be $\sim10\%$ \citep[e.g.,][]{Bland-Hawthorn2016TheProperties}.

Galactic spectroscopic surveys such as Gaia-ESO \citep{Gilmore2012TheSurvey}, SEGUE \citep{Yanny2009SEGUE:14-20}, GALAH \citep{DeSilva2015TheMotivation}, and APOGEE \citep{Majewski2017TheAPOGEE} have provided the metallicity information for a large number of disk stars.
The vertical gradient of alpha abundance and metallicity \citep[e.g.,][]{Hayden2014ChemicalDisk} have indicated that the thick disk consists of alpha-enhanced and metal-poorer populations \citep[e.g.,][]{Bland-Hawthorn2016TheProperties}. 
Indeed, in many studies, two chemically-distinct sequences are thought to be separated in the plane of [Fe/H] and [$\alpha$/Fe], which further supported the idea of two-component disks 
\citep[e.g.,][]{Lee2011FormationSample, Anders2014ChemodynamicsData, Bensby2014ExploringNeighbourhood, Duong2018TheNeighbourhood, Mackereth2019DynamicalitGaia}.
These two sequences exhibit different spatial distributions and kinematic properties; the alpha-rich metal-poor sequence tends to have a shorter scale length, a larger scale height, and a slower rotational velocity than the alpha-poor sequence. 
On the other hand, some studies found that disk stars of varying chemical abundance show gradually varying thickness, arguing against the presence of two distinct disk structures \citep[e.g.,][]{Bovy2012TheDisk}.
Therefore, it remains uncertain whether the thick disk is truly a distinct component with a different origin.

Thick disk studies have been conducted on other galaxies as well.
Due to observational limitations, it has been possible only in a few nearby edge-on spiral galaxies to resolve individual disk stars and measure the vertical number density \citep[e.g.,][]{Seth2005Population}. 
Hence, most studies were based on the vertical brightness profile. 
For example, \cite{Yoachim2006StructuralGalaxies} have measured the vertical profiles of 34  edge-on disk galaxies with a wide range of stellar mass. 
They found that the luminosity ratio between the thin and thick disks, $L_{\rm thick}/L_{\rm thin}$, depends on the galaxy stellar mass: thin disks contribute more to the total disk luminosity in more massive galaxies, while the luminosity of thick disks is almost comparable to that of the thin disk in low-mass galaxies \citep[see also][]{Comeron2011ThickBaryons, Comeron2014EvidenceImaging, Martinez-Lombilla2019}. 
The kinematics and stellar population properties of the two disks have been studied only for a small number of galaxies through spectroscopy of the thin disk (close to the galactic midplane) and thick disk regions (typically a few kpcs away from the midplane) \citep[e.g.,][]{Yoachim2008LickGalaxies, Comeron2019, Pinna2019TheEnvironment, Kasparova2020ANGC7572}.
They have confirmed that thick disks are, in general, older, relatively metal ([Fe/H])-poorer, and alpha-enhanced.

Theoretically, several scenarios have been proposed to illustrate the formation of thick disks. One of the scenarios is the kinematic heating (thickening) of a preexisting thin disk, caused by several possible sources, including minor mergers \citep[e.g.,][]{Quinn1993HeatingMergers, Kazantzidis2008ColdAccretion,Villalobos2008SimulationsDiscs}, spiral and barred structures \citep[e.g.,][]{SellwoodJ.A.1984SpiralFormation,Saha2010TheGalaxies, Grand2016VerticalContext}, giant molecular clouds \citep[e.g.,][]{Spitzer1951TheVelocities, Aumer2016Age-velocityGalaxies}, and stellar clumps \citep[e.g.,][]{Bournaud2009TheRedshift,Silva2020} 
Also, orbital diffusion (radial migration) has been suggested as possible mechanism to form a thick disk 
\citep[e.g.,][]{Roskar2008BeyondGrowth:,Schonrich2009OriginDiscs, Loebman2011TheMigration, Fouvry2017, Halle2018}; stars migrating outward due to the interaction with spiral arms also drift further away from the midplane as well. 
Recently, \cite{Sharma2020} showed from their chemodynamical model of MW that radial migration is essential for forming both alpha-enhanced and alpha-poor sequences. They concluded that their model does not require ``distinct'' thick component, as both sequences are formed in a ``continuous'' star formation and evolution history of the MW.

In addition to triggering disk heating \citep[e.g.,][]{Helmi2018TheDisk}, mergers are also believed to play an important role in forming thick disks in several ways.
For example, \cite{Abadi2003SimulationsDisks} suggested, based on a simulated galaxy, that the thick disk could be formed by accretion of the stars (formed ex situ) from disrupted satellite galaxies 
\citep[see also][]{Gilmore2002DecipheringWay, Wyse2006FurtherGalaxies}. 
Also, several studies using simulations have claimed that gas-rich mergers can induce starbursts in the main galaxies and that many of the thick disk stars are formed in situ during these chaotic merger events \citep[e.g.,][]{Brook2004TheUniverse, Brook2012ThinGalaxy, Grand2020SausageMerger}.
Based on four MW-mass simulated galaxies, \cite{Buck2020OnMigration} supported the idea that gas-rich mergers result in chemical bimodality of disk stars, but they found that low-alpha sequence is formed as a result of meteal-poor gas-rich mergers that dilute the interstellar medium \citep[see also][]{Bonaca2020}. 
Most recently, \cite{Agertz2020VINTERGATANGalaxy} found in their one MW-mass simulated galaxy that a alpha-poor sequence started forming after the last major merger and the subsequent formation of an outer metal-poor gas disk \citep[see also][]{Renaud2020VINTERGATANMergers}.
Thus, they concluded that the two alpha sequences (alpha-enhanced/ alpha-poor) are formed during different (merger-dominated/ quiescent) growth phases of the galaxy, and thus, this transition of the growth phases is the key to creating the bimodal distribution of the stars in the chemical plane.
Contrary to this sequential formation of the thin and thick disks, however, some studies proposed a co-formation scenario of the thin and thick disks. For example, based on the kinematic and chemical properties of old disk stars in the solar neighborhood, \cite{Silva2020} claimed that both thin and thick disks started to form early, which can be explained by a scenario in which a thick disk is formed by scattering by clumps.

Recent studies using cosmological hydrodynamic simulations have further elaborated on the origin of the thin and thick disks.
The overall picture on disk evolution described by many cosmological galaxy simulations is the ``inside-out'' and ``upside-down'' formation; when stellar particles in the simulated galaxies are divided into co-eval populations, older cohorts have larger scale heights and shorter scale lengths \citep[e.g.,][]{Stinson2013MaGICCWay, Martig2014DissectingPopulations,Minchev2015OnDisks}.
In this framework, the two-stage formation of the thick disk is expected 
\citep[e.g.,][]{Ma2017TheGalaxy, Buck2020NIHAO-UHD:Simulations}; stars formed earlier are born with thicker distributions \citep[e.g.,][]{Bird2013InsidePopulations, Grand2016VerticalContext, Bird2020}, as they are formed out of more turbulent gas at higher redshifts \citep[e.g.,][]{Bournaud2009TheRedshift, Forbes2012EvolvingFormation}. 
Subsequently, gradual kinematic heating of the later-formed thinner disk contributes to the growth of the thick disk \citep[e.g.,][]{Martig2014DissectingRelation}. 
Based on the vertical profiles of the mono-age stellar populations in the simulated galaxies, \cite{Buck2020NIHAO-UHD:Simulations} showed that the thick disk is not a distinct component, and rather, the double-component vertical structure can be understood as a gradually-varying mixture of young and old stars with vertical height from the galactic plane.

This study aims to explore whether the spatially-defined thin and thick disks are 
distinct components formed by different mechanisms. 
We use 18 disk galaxies from the {\nh} simulation that reached $z=0.3$ \citep{Dubois2020} and one disk galaxy at $z=0$ from the  {\gal} simulation (Peirani et al. in prep). 
These simulations have relatively high spatial and mass resolutions compared to other present-day simulations and thus are ideal to study the disk structure of galaxies.
Furthermore, the up-to-date baryonic physical prescriptions including the supernovae (SN) and active galactic nuclei (AGN) feedback have been implemented in both simulations. 
Since these simulations did not trace the alpha element separately from the total metallicity, this study focuses on the {\em spatially-defined} thin and thick disks.

The paper is organized as follows; Section~\ref{sec:methodology} describes the two simulations and the sample galaxies used in this study.
In Section~\ref{sec:relations_from_fit}, we first apply the conventionally-accepted double-component fitting to the vertical profiles of galaxies, and check if the simulations reproduce the observed properties of the two disks. 
In Section~\ref{sec:decomposition}, we separate the thin and thick disks geometrically based on the vertical profiles and compare them in terms of ex-situ contribution, age, metallicity (Section~\ref{sec: decomposition_age_metal}), kinematics (\ref{sec: decomposition_kinematics}), and the properties at birth (\ref{sec: decomposition_birthprops}).  
We divide in Section~\ref{sec:evol} the stellar particles formed in situ into mono-age groups and trace the vertical distribution of each mono-age group with time in order to understand how the two disks can be interpreted in terms of disk evolution. 
In Section~\ref{sec:conclusion}, we conclude that the two disks are not entirely distinct in terms of formation mechanisms; instead, they are the results of how galactic disks develop and evolve with time.


\section{Methodology} \label{sec:methodology}
\subsection{Simulations - NewHorizon and Galactica}
The {\nh} \citep[][hereafter NH]{Dubois2020} and {\gal} (Peirani et al. in prep) simulations are both high-resolution cosmological hydrodynamic zoom-in simulations from the larger volume of the Horizon-AGN simulation \citep{Dubois2014DancingWeb}.
These simulations are run with {\sc ramses} \citep{Teyssier2002CosmologicalRefinement}, an Eulerian hydrodynamics code with adaptive mesh refinement (AMR), and share the same cosmology based on the WMAP-7 data \citep{Komatsu2011SEVEN-YEARINTERPRETATION}: 
Hubble constant $H_{0}=70.4\,{\rm km\,s^{-1}\,Mpc^{-1}}$, 
total baryon density $\Omega_{b}=0.0455$, 
total mass density $\Omega_{m}=0.272$,
dark energy density $\Omega_{\Lambda}=0.728$, 
amplitude of power spectrum $\sigma_{8}=0.809$, 
and power spectral index $n_{s}=0.967$.
The parent simulation, Horizon-AGN is a large-volume, $(100\,\rm Mpc/h)^3$, cosmological simulation with a spatial resolution of $\Delta x\gtrsim 1\,\rm kpc$, and the mass resolution of $m_{\rm DM}=8\times10^7\,\msun$ (a dark matter particle) and $m_{\rm stellar}=2\times10^6\,\msun$ (a stellar particle).

{\nh} zooms in a spherical region with a radius of $\rm 10\,comoving\,Mpc$ in the field environment of Horizon-AGN.
The refinement of each cell in the zoomed region is allowed when the mass of the cell is greater than 8 times the initial mass resolution, or the size of a dense cell $(n_{\rm H}> 5\,\rm cm^{-3})$ is smaller than Jeans length, to reach $\Delta x\sim34\,\rm pc$ at $z=0.0$.
Additional global cell refinement occurs at $a_{\rm exp}$ $\simeq$ $0.1,\,0.2,\,0.4,$ and $0.8$ to keep the minimum cell size roughly constant.  
In this study, we used galaxies at $z=0.3$, which is close enough to compare the properties of galaxies with those in the local Universe.
The dark matter and stellar particle resolutions are $m_{\rm DM}=10^6\,\msun$ and $m_{\rm stellar}=10^4\,\msun$, respectively, and the spatial resolution is $\Delta x\gtrsim 52\,\rm pc$ (in physical scale) at $z=0.3$ \footnote{The simulation is currently running and $z=0.3$ is the lowest available redshift suitable for this analysis.}.
The high-resolution nature of the NH simulation was found to be very useful to pin down the origins of the disk and spheroidal components of field galaxies \citep[e.g.,][]{Park2019}.

The ongoing {\gal} project focuses on individual galaxies in the field environments, extracted from the Horizon-AGN volume (yet not in the {\nh} volume) to explore their evolution in detail down to $z=0.0$.
The spatial resolution at the final epoch ($z=0.0$) is $\Delta x\gtrsim 34\,\rm pc$ (in physical scale), as targeted by {\nh}.
Since all NH galaxies can only be studied down to $z\sim0.3$, we also explored one {\gal} galaxy further down to $z=0.0$ in more detail. 
Here we briefly describe the baryonic prescriptions implemented in both {\nh} and {\gal} simulations. 
A more detailed description can be found in the introduction paper of the {\nh} project \citep{Dubois2020}.

Gas can cool down to $0.1\,\rm K$ through primordial and metal cooling following the model of \cite{Sutherland1993CoolingPlasmas} (down to $10^{4}\,\rm K$) and \cite{Dalgarno1972HeatingRegions} (below $10^{4}\,\rm K$).  
After the reionization epoch at $z=10$, gas is heated by a uniform UV background radiation based on \cite{Haardt1996RadiativeBackground}, and the heating rates are assumed to be reduced by ${\rm exp}(-n/n_{\rm shield})$ in optically thick regions (gas density, $n_{\rm H}>n_{\rm shield}=0.01\,\rm cm^{-3}$) due to self-shielding.
Stars can form in a cell where the hydrogen number density is greater than $n_{\rm H}=\rm 10\,cm^{-3}$ and temperature is lower than $2\times10^{4}\,\rm K$, based on the Schmidt relation \citep{Schmidt1959TheFormation}: ${\dot \rho} = \epsilon_{*}\,\rho_{g}/t_{\rm ff}$, where ${\dot \rho}$ is the star formation rate, $\epsilon_{*}$ is the star formation efficiency, $\rho_{g}$ is the gas density, and  $t_{\rm ff}$ is the free-fall time. 
The star formation efficiency ($\epsilon_{*}$) varies depending on the local turbulent Mach number and the virial parameter, following \cite{Kimm2017Feedback-regulatedReionisation}.

Each stellar particle with a mass of $10^4\,\msun$ is assumed to be a simple stellar populations following the Chabrier initial mass function \citep{Chabrier2005The2005}, with a lower and upper mass cutoff of $0.1\,\msun$ and $150\,\msun$.
We assume that the minimum mass of a star for the Type II supernova (SN) explosion is $6\,\msun$. 
When a stellar particle becomes older than $5\,\rm Myr$, SN explosion is assumed to take place, returning $31\%$ of the mass of the stellar particle to the surroundings, with the metal yield of 0.05. 
Mechanical SN feedback scheme is employed following \cite{Kimm2014EscapeStars}.

Supermassive black holes (BH) are represented as sink particles placed in cells where gas and stellar densities exceed the threshold of star formation and stellar velocity dispersion higher than $20\,\rm km/s$.  
New sink particles are not allowed to form within a distance of $\rm 50\,comoving\,kpc$ from preexisting sink particles, and two existing BHs are assumed to merge with each other when they get within $150\,\rm pc$. 
The initial seed mass of a BH is $10^{4}\,\msun$ and it grows with a Bondi-Hoyle-Lyttleton accretion rate \citep{Hoyle1939TheVariation, Bondi1944OnStars}: $\dot M_{\rm BH} = (1-\epsilon_{r}\,\dot M_{\rm Bondi})$ where $\dot M_{\rm Bondi} = 4 \pi \rho (G M_{\rm BH})^2/(u^2 + c_s^2)^{3/2}$, $u$ is the average velocity of BH relative to gas's velocity, $c_{s}$ is the average sound speed, $\rho$ is the average gas density, and $\epsilon_{r}$ is the spin-dependent radiative efficiency. 
The maximum limit of the accretion rate is set to the Eddington rate. 
Two different feedbacks from AGN are modeled based on the ratio of the gas accretion rate to the Eddington rate ($\chi$), following \cite{Dubois2012Self-regulatedSimulations}; the radio mode ($\chi<0.01$) and the quasar mode ($\chi>0.01$). 
In the radio mode, AGN releases mass, momentum, and energy in the form of bipolar jets with a spin-dependent feedback efficiency, while in the quasar mode, it only deposits thermal energy  isotropically into the surrounding gas cells (i.e., increases temperature).

\subsection{Sample selection}
\begin{figure*}
    \centering
    \linespread{1.0}\selectfont{}
    \includegraphics[width=\textwidth]{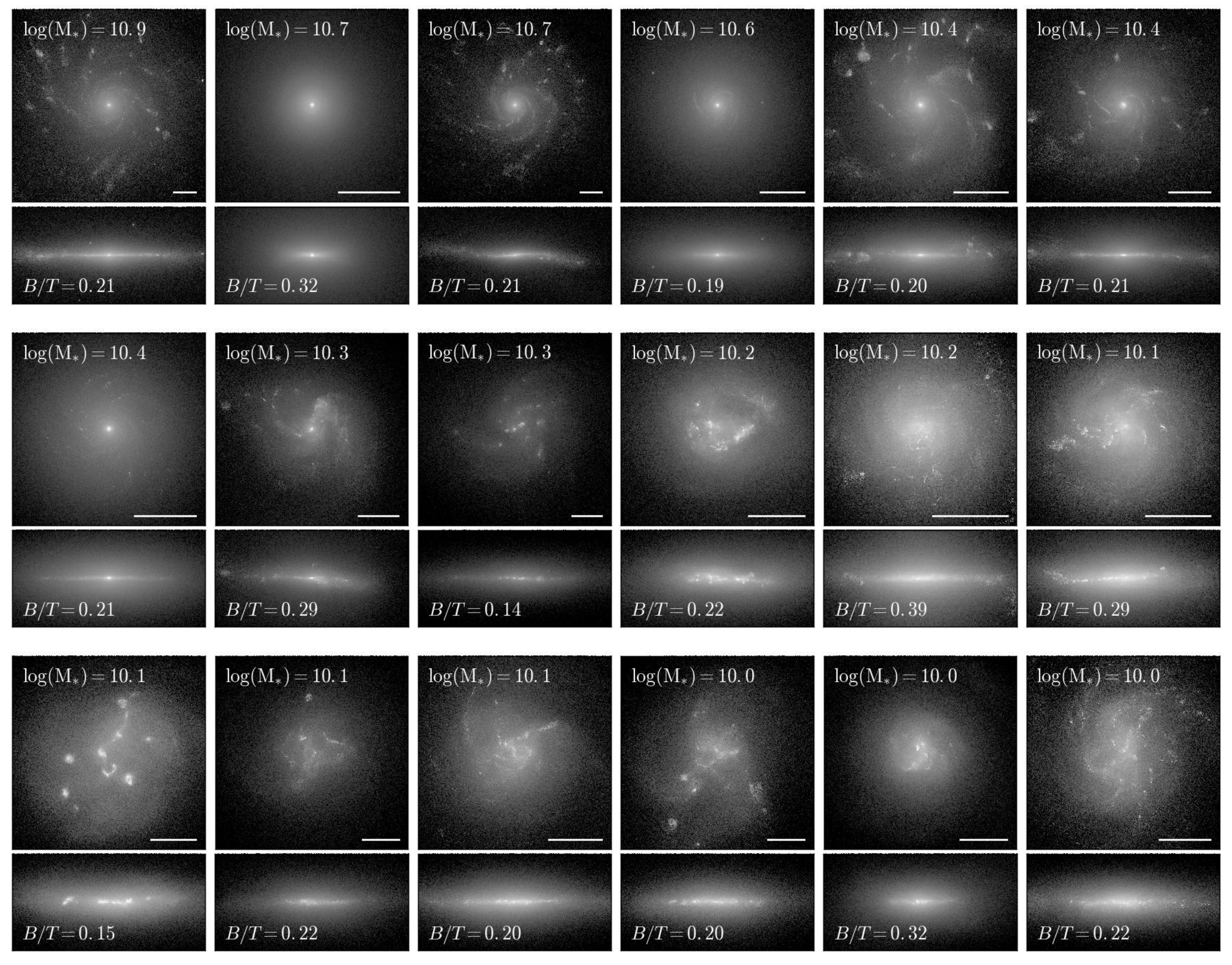}
    \caption{The $\sl r$-band face-on and edge-on images of the 18 selected NH galaxies at $z=0.3$, in descending order of stellar mass of galaxies. The extent of each box for the face-on image is $2\,R_{90}$ of each galaxy, while the height of the box for the edge-on image is $R_{90}$. The $R_{90}$ is the 3D radius containing 90\% of the total stellar mass of a galaxy. 
    The white bar represents $\rm 5\,kpc$. The kinematic bulge-to-total ratio ($B/T$) of each galaxy is presented at the bottom.}
    \label{fig:nh_diskgal_table}
\end{figure*}

Galaxies in the {\nh} simulation are identified based on the AdaptaHOP algorithm \citep{Aubert2004TheHaloes} with the most massive sub-node mode \citep{Tweed2009BuildingSimulations}, and a minimum number of stellar particles required for a galaxy is 50.
At $z=0.3$, there are 54 identified galaxies with a stellar-mass larger than $10^{10}\msun$.
We apply the stringent mass cut despite the high mass resolution in order to secure a large number of star particles for reliable statistical analysis.
Since the zoomed region of the {\nh} simulation is embedded in its parent simulation, Horizon-AGN, some halos in the outskirts of the zoomed region are contaminated by low-resolution DM particles. 
Hence, we only selected the galaxies in the halos with contamination fraction lower than 0.001 (0.1\%), which gives us 31 galaxies with masses above $10^{10}\msun$ (22 of them are contamination-free). 
In this study, we have selected 18 disk galaxies from 31 galaxies that are visually undisturbed and have bulge-to-total ratios, $B/T \leq 0.4$. 
The bulge-to-total ratio is measured using the stellar particles inside $R_{90}$ of each galaxy, the radius within which 90\% of the total stellar mass is contained.
The ``bulge'' component is defined kinematically using the orbital circularity parameter, $\epsilon$ \citep{Abadi2003SimulationsDisks}, which quantifies how close a stellar orbit is to the circular orbit in the galactic plane by comparing the angular momentum of the stellar particle in the net rotating direction (z-axis) of a galaxy ($J_{\rm z}$) and that of a circular orbit with the same energy ($J_{\rm circ}$): $\epsilon=J_{\rm z}/J_{\rm circ}$.
The bulge mass is measured as twice the mass sum of the stellar particles with $-0.5<\epsilon<0$, assuming that the circularity of the bulge stars is symmetrically distributed around 0 (i.e., random-dominated component). 
Thus, $B/T$ gives the mass fraction of ``non-disk'' components with dispersion-dominated kinematics, including the stellar halo. 
Figure~\ref{fig:nh_diskgal_table} shows the $\sl r$-band face-on and edge-on images of the 18 selected NH galaxies at $z=0.3$.
The properties of the simulated galaxies used in this study are given in Table~\ref{table:scale_height_and_luminosity_ratios}.

\section{Properties of thin and thick disks} 
\label{sec:prop}
First, we will check whether the models in the simulations reproduce the observed characteristics of the thin and thick disks, such as scale heights and luminosity ratios, when the double-component fitting is applied to the vertical profiles of the simulated galaxies.
Then, we will separate the two disk components based on the vertical profile and identify which stellar particles belong to each component. 
In an attempt to see how physically distinct the two components are, we will focus on the difference between the two components in various properties, such as the fraction of stars formed ex situ, age, metallicity, kinematics, and birthplaces.

\subsection{The two-component fits to the vertical profiles}
\label{sec:relations_from_fit}
\subsubsection{The Galactica galaxy}
\begin{figure}
    \linespread{1.0}\selectfont{}
    \includegraphics[width=\columnwidth]{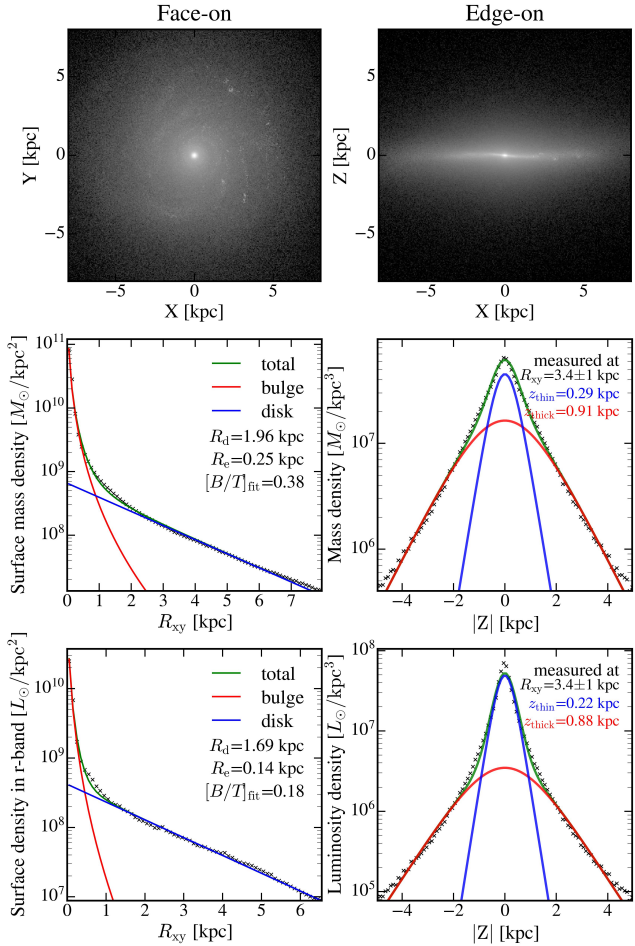}
    \caption{(Top) The face-on and edge-on $\sl r$-band images of the {\gal} galaxy and the radial and vertical profiles measured in mass (middle) and $\sl r$-band (bottom). The radial profiles are fitted with the combination of Sersic and exponential disk profile, returning the scale length of the disk ($R_{\rm d}$) and bulge-to-total ratios, $[B/T]_{\rm fit}$. The two-component profile is applied to the mass/$\sl r-$band vertical profile of the galaxy, from which the scale heights of the thin and thick disks are measured.}
    \label{fig:image_profile_Galactica}
\end{figure}

First, we check the radial and vertical profile of the {\gal} galaxy at $z=0.0$.
The top panels of Figure~\ref{fig:image_profile_Galactica} show the face-on and edge-on $\sl r$-band images of the {\gal} galaxy. 
The $\sl r$-band flux of each stellar particle is calculated based on its age and metallicity following the \cite{Bruzual2003Stellar2003} stellar population model, without dust extinction. 
The galaxy's stellar mass and size are about half that of the MW; the stellar mass of this galaxy is $2.75\times10^{10}\,\msun$, and the half-mass radius ($R_{50}$) and $R_{90}$, the radius within which 90\% of the total stellar mass is contained, are $\rm 1.9\,kpc$ and $\rm 8.0\,kpc$, respectively.

We measure the radial and vertical profiles of this galaxy in both mass and $\sl r$-band luminosity.
The radial profile is measured from the face-on images and the combination of the Sersic \citep{Sersic1963InfluenceGalaxy} and exponential disk profiles is applied to the radial profile, which gives the bulge-to-total ratios, $[B/T]_{\rm fit}$, of 0.38 (mass) and 0.18 ($\sl r$-band).
To measure the vertical distribution, we use the cylindrical coordinate where z-axis is the direction of the net angular momentum of the stellar particles inside $R_{90}$. 
The vertical distribution is measured in a cylindrical bin where $R_{\rm xy}=\rm 3.4\pm 1\,kpc$, corresponding to $2\,R_{\rm d}$, where $R_{\rm d}$ is disk scale length measured in $\sl r$-band, and fitted with a double-component profile \citep[e.g.,][]{Kruit1981SurfaceII,  Yoachim2006StructuralGalaxies, Comeron2011ThickBaryons}: 
\begin{equation}
    \hspace{-0.05cm} \rho\,(\rm Z ) = \rho_{thin}\,\rm sech^2\!\left( \frac{|Z|}{2\,{\it z}_{thin}} \right)
    \!+\! \rho_{thick}\,\rm sech^2\!\left( \frac{|Z|}{2\,{\it z}_{thick}} \right)
\end{equation}
where $\rho_{\rm thin}$ and $\rho_{\rm thick}$ are densities of the thin and thick disks in the galactic midplane ($\rm |Z|$=0), and $z_{\rm thin}$ and $z_{\rm thick}$ are their scale heights. The scale heights of the thin and thick disks are $z_{\rm thin}=290\,\rm pc$ and $z_{\rm thick}=910\,\rm pc$ in mass, while, in $\sl r$-band, both disks have slightly shorter scale heights of $z_{\rm thin}=220\,\rm pc$ and $z_{\rm thick}=880\,\rm pc$.

\begin{figure*}
    \linespread{1.0}\selectfont{}
    \includegraphics[width=\textwidth]{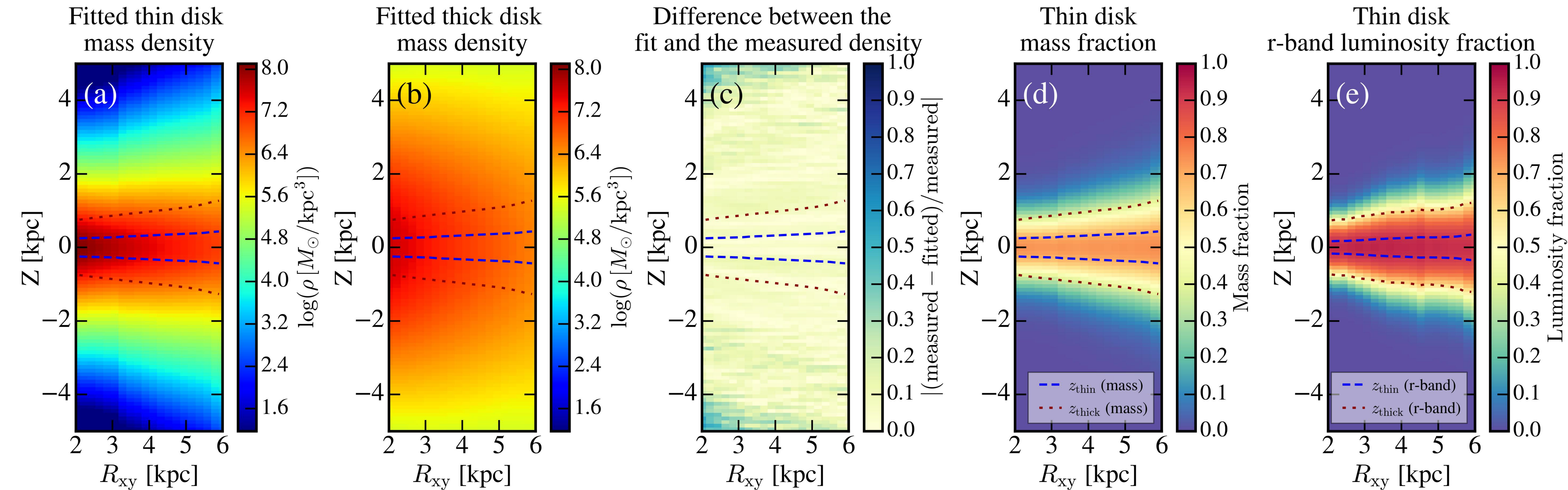}
    \caption{The projected map of the {\gal} galaxy showing the mass density of the (a) thin and (b) thick disks derived from the double-component fit, in the cylindrical regions where $2\,\rm kpc<\textit{R}_{xy}<6\,\rm kpc$ and $\rm |Z|<5\,kpc$. (c) The difference between the fitted density derived by the double-component fit and the calculated mass density. (d) The mass ratios of the fitted thin disk. (e) The $\sl r$-band luminosity ratios of the fitted thin disk. Only the two components, thin and thick disks are assumed in this region, so the sum of the ratios of the thin and thick disks is set to 1. The blue dashed and red dotted lines in each panel are the scale heights of the thin and thick disks in mass (in Panel a to d) and in $\sl r$-band (in Panel e).} 
    \label{fig:projected_map}
\end{figure*}

To visualize the relative importance of the thin and thick disks in different galactic regions, we measure the vertical distributions in the cylindrical region with radii ($R_{\rm xy}$) of $\rm 2-6\,kpc$ with a bin size of $\rm 0.2\,kpc$ and plot the fitted density of the thin and thick disks in each region. 
In each radial bin, the vertical distribution is obtained by measuring the number (mass) density of the stellar particles inside the window of $\rm \pm 0.5\,kpc$, to avoid high fluctuations due to the small number of stellar particles. 
In Figure~\ref{fig:projected_map} (a) and (b), it is clear that the thin disk component is concentrated near the galactic midplane, while the thick disk component smoothly spreads over the disk regions ($R_{\rm xy}=\rm 2-6\,kpc$ and $\rm |Z|<5\,kpc$). 
The double-component fit seems reasonably good (at least up to $\rm |Z|\sim4\,kpc$), as can be seen in Panel (c) where it shows the difference between the sum of the fitted densities of the thin and thick disks and the calculated mass density in each region.

In Figure~\ref{fig:projected_map} (d), we also construct the projected map showing the thin disk mass ratios in this region. 
Since we have assumed only two components, the sum of the thin and thick disk mass ratios in each bin is set to 1. 
The vertical height where the mass densities of the thin and thick disks are comparable, represented as the light-yellow regions, increases with the projected radius as well, as a result of disk flaring.
The scale heights of thin and thick disks, represented as blue dashed and red dotted lines, both increase with the projected radius at which the vertical profile is measured; from $R_{\rm xy}=2\,\rm kpc$ to $\rm 6\,kpc$, thin disk increases from $\rm 260\,pc$ to $\rm 490\,pc$, and thick disk increases from $\rm 770\,pc$ to $\rm 1400\,pc$ (by a factor of $\sim2$).
This flaring of the disk can also be found in other galaxy simulations, while the degree of flaring varies across simulations \citep[e.g.,][]{Minchev2015OnDisks, Grand2017TheN, Ma2017TheGalaxy, Buck2020NIHAO-UHD:Simulations}, which may be due to the different merger histories of the galaxies, given that flaring is inevitably affected by mergers \citep[e.g.,][]{Bournaud2009TheRedshift}.

Figure~\ref{fig:projected_map} (e) presents the thin disk $\sl r$-band luminosity ratios in the same region, derived from the vertical profiles in $\sl r$-band in the same way as the mass ratios. 
The flaring is also visible in the $\sl r$-band profile, nearly to the similar degree as the flaring seen in the mass profile.
The $\sl r$-band luminosity near the galactic midplane, however, is more dominated by the thin disk ($>80\%$), as young stars in the thin disk contribute significantly to the total luminosity, which makes the thin disk component more pronounced with a stronger slope change in the vertical distribution (Figure~\ref{fig:image_profile_Galactica}).

According to Panels (d) and (e), which are based on the vertical profiles, $\sim$30\% of the mass and $\sim$10\% of the $\sl r$-band luminosity near the galactic midplane appear to belong to the thick disk component.  
These values are broadly consistent with other simulations.
For example, \cite{Ma2017TheGalaxy} mentioned that $\approx36\%$ of the stellar mass at the solar radius ($R_{\rm xy}=8\,\rm kpc$) in their simulated MW-mass galaxy belongs to the thick disk according to the two-component fit to the vertical profile. 
In \cite{Brook2012ThinGalaxy}, they identified the thick disk component in the chemical plane ($\rm [\alpha/Fe]-[Fe/H]$), and it contributed to $\approx 27\%$ of the stellar mass in the solar neighborhood.  
On the other hand, several studies performed the kinematic decomposition for all stellar particles in the simulated galaxies and showed that the mass of the thick disk is nearly comparable to that of the thin disk \citep[e.g.,][]{Abadi2003SimulationsDisks,Obreja2018IntroducingGalaxy,Obreja2019NIHAOHaloes}. 
The mass or luminosity ratio between the thin and thick disks seems to vary greatly depending on the measuring range and methods (e.g., derivation from the decomposition based on kinematics or chemical abundance).

\subsubsection{The NH sample galaxies at $z=0.3$}
\begin{figure*}
    \linespread{1.0}\selectfont{}
    \includegraphics[width=\textwidth]{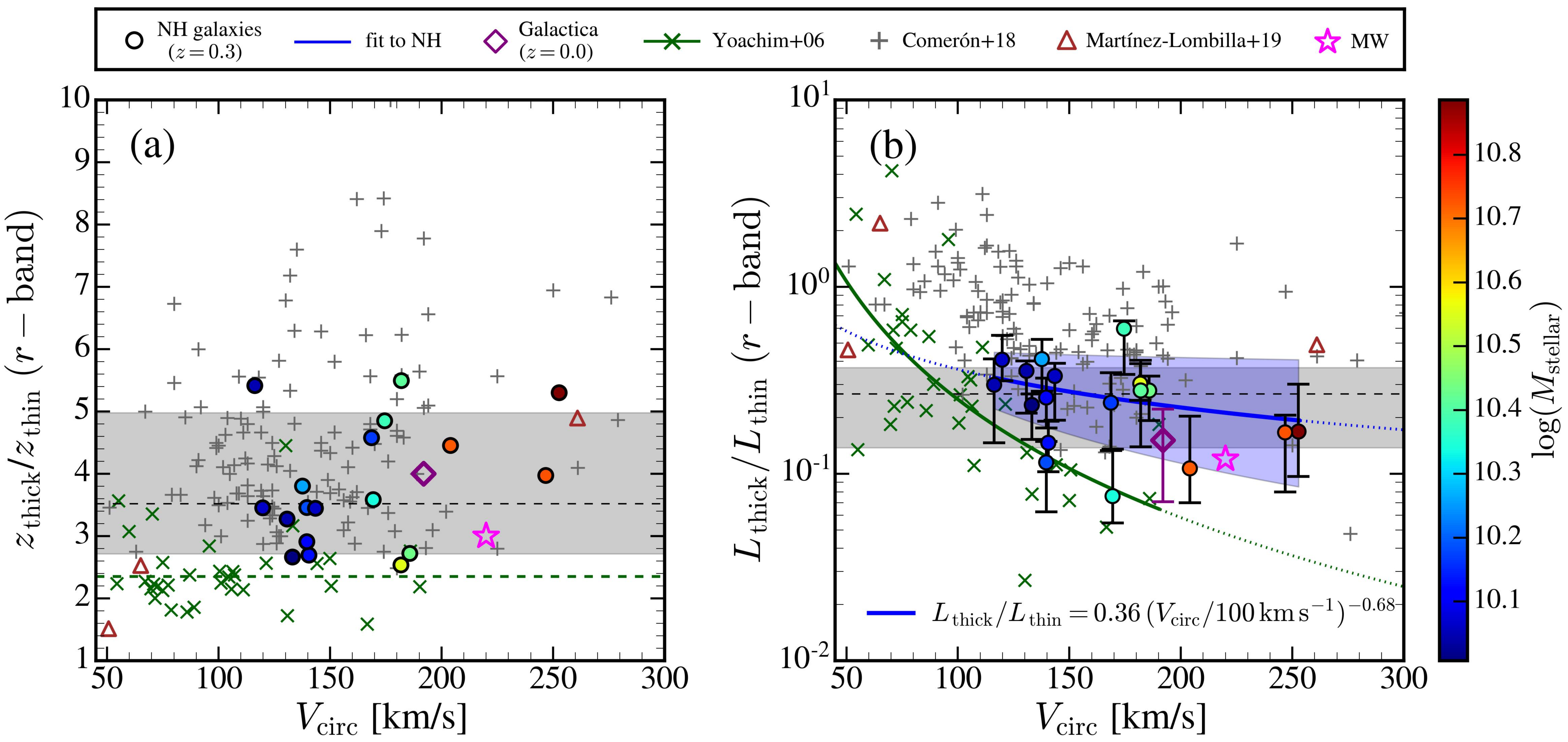}
    \caption{(a) The scale height ratio of the thin and thick disks ($z_{\rm thick}/z_{\rm thin}$) derived from $\sl r$-band vertical profile as a function of the circular velocity ($V_{\rm circ}$) of the NH galaxies. (b) The $\sl r$-band luminosity ratio of the two disks ($L_{\rm thick}/L_{\rm thin}$) in the regions near the galactic midplane ($\rm |Z|<1\,kpc$). The error bars cover a range of luminosity ratios measured from the midplane at $\rm |Z|=0$ (lowest) to the ratios measured up to $\rm |Z|<2\,{\it z}_{\rm thick}$ (highest). The color represents the stellar mass of the NH galaxies at $z=0.3$. The values of the scale height and luminosity ratios between the two disks in each galaxy are listed in Table~\ref{table:scale_height_and_luminosity_ratios}. The dashed gray horizontal line in each panel shows the median value of the scale height/luminosity ratios of the NH galaxies with a gray shade showing the range from 16th to 84th percentile. 
    The values of the {\gal} galaxy at $z=0.0$ are presented as open purple diamonds with $z_{\rm thick}/z_{\rm thin}=4.06$ and $L_{\rm thick}/L_{\rm thin}=0.15$, while the MW vales are marked as open magenta stars: $z_{\rm thick}/z_{\rm thin}\sim3.0$ \citep[e.g.,][]{Bland-Hawthorn2016TheProperties} and $\rho_{\rm thick}/\rho_{\rm thin}\sim0.12$ \citep[e.g.,][]{Juric2008TheDistribution}.
    The gray ``$+$'' markers are the values of observed galaxies from \cite{Comeron2018} and the brown triangles are from \cite{Martinez-Lombilla2019}. Also, note that in Panel (b), the values of these observed galaxies are mass ratios between the two disks, $M_{\rm thick}/M_{\rm thin}$, instead of luminosity ratios.
    The green ``$\times$'' markers are the values of observed galaxies from \cite{Yoachim2006StructuralGalaxies}, and the dashed green line in Panel (a) is the median scale height ratios of their sample galaxies.
    The green solid line is the observed relation suggested by \cite{Yoachim2006StructuralGalaxies} obtained by fitting a power law to their samples (with $V_{\rm circ}\sim 50-190\,\rm km/s$): $L_{\rm thick}/L_{\rm thin}=0.25\,(V_{\rm circ}/100\,\rm km\,s^{-1})^{-2.1}$. 
    The blue solid line represents the fitted power law to the NH galaxies (with $V_{\rm circ}\sim 110-260\,\rm km/s$), and a shade indicates $1\,\sigma$ of the fit: $L_{\rm thick}/L_{\rm thin}=0.36\,(V_{\rm circ}/100\,\rm km\,s^{-1})^{-0.68}$. 
    We find that thin and thick disks are reproduced in our simulations when the two-component fits are applied and the resulting scale heights and luminosity ratios between the two disks are broadly consistent with observations; there is a slight tendency in the NH galaxies that the contribution of the thick disk to the total luminosity decreases with galaxy's stellar mass.}
    \label{fig:scaleheight_density_ratios}
\end{figure*}

We also applied the same two-component fits to the $\sl r$-band vertical profiles of the 18 NH sample galaxies at $z=0.3$. 
Note that the vertical profile is measured at $2\,R_{\rm d}$ of each galaxy to account for the different disk sizes of the galaxies. 
Figure~\ref{fig:scaleheight_density_ratios} (a) shows the scale height ratios of the thin and thick disks measured in $\sl r$-band as a function of the circular velocity ($V_{\rm circ}$) of the galaxies (See also Table~\ref{table:scale_height_and_luminosity_ratios} for the properties of each galaxy). 
The circular velocity is measured at $R_{90}$, following $V_{\rm circ} = \sqrt{GM(<r)/r}$. 
The colors inside the circles represent the stellar mass of the NH galaxies at $z=0.3$, indicating that more massive galaxies have higher circular velocities. 
We also include the values of the {\gal} galaxy at $z=0$, with $z_{\rm thick}/z_{\rm thin}=4.06$ in $\sl r$-band, as an open diamond.
The magenta star shows the scale height ratio of the MW, which is about 3 \citep[e.g.,][]{Gilmore1983GalacticDisc,Juric2008TheDistribution}. 
Note that the value of the MW is measured based on the number density, not the photometric profile, yet it is still in a good agreement with the NH and {\gal} galaxies.

Our sample of the NH galaxies shows a broad range of scale height ratios in $\sl r$-band, broadly consistent with observations \citep[e.g.,][]{Yoachim2006StructuralGalaxies, Comeron2018, Martinez-Lombilla2019}.
The diagram is dominated by scatter both in observational and simulation data, but there is a hint of a positive trend in the sense that a more massive galaxy shows a larger ratio between the vertical scale height of thick and thin disks. The linear fit to the observed data, for reference, has a slope of 0.013 and one standard deviation error is 0.0026.
The median value across the sample of the galaxies is $z_{\rm thick}/z_{\rm thin}=3.52$, represented as a dashed line, and a gray shade shows the 16th to 84th percentiles.

Figure~\ref{fig:scaleheight_density_ratios}(b) shows the $\sl r$-band luminosity ratio between the thick and the thin disks ($L_{\rm thick}/L_{\rm thin}$), as a function of the circular velocity of the galaxy.
We plotted the luminosity ratios in the regions near the galactic midplane ($\rm |Z|<1\,\rm kpc$) with error bars covering a range of the ratios measured in the vertical range of $\rm |Z|<2\,z_{\rm thick}$.
The colors inside the circles again represent the stellar mass of the galaxies. 
The dashed line shows the median of the ratios in the NH galaxies, and the gray shade, again, represents the range of 16th to 84th percentiles of the luminosity ratios in the samples.

We found that the $\sl r$-band luminosity ratios seem to decrease slightly with the $V_{\rm circ}$ of the galaxies. 
A similar trend was found in \cite{Yoachim2006StructuralGalaxies} by fitting a power law to their samples, which is represented as a green solid line in Panel (b).
We also applied the power-law fit to the luminosity ratios of the NH galaxies and the blue line shows the obtained fit with a shade of $1\,\sigma$: $L_{\rm thick}/L_{\rm thin}=0.36\,(V_{\rm circ}/100\,\rm km\,s^{-1})^{-0.68}$.
It seems that our trend is shallower than the observed one, indicating that thick disks of our simulated galaxies may be brighter than the observations. 
The $L_{\rm thick}/L_{\rm thin}$ of {\gal} (the purple diamond) and the MW\footnote{The value of the MW was taken as the local number density between the thin and thick disks: $\rho_{\rm thick}/\rho_{\rm thin}=0.12$ \citep{Juric2008TheDistribution}} (the magenta star) fall within the $1\,\sigma$ of our trend.

Given that MW has been forming stars at $\sim 1\,\msun/\rm yr$ \citep[see e.g.,][]{Robitaille2010}, the estimated growth of the disk during the last $3.5\,\rm Gyr$ would be only by a factor of $5\sim10\%$\footnote{This value is a rough estimate based on the current star formation rate (SFR) of the MW. Taking into account the SFR over the past 3 Gyrs, the estimate could be higher by a factor of 2 or so.}. Therefore, we expect that the properties of the thin and thick disks we examined for the NH galaxies at $z=0.3$ would be roughly similar even at $z=0.0$, unless they experience mergers after $z=0.3$ that could severely disturb the disk structures. As a check, we explored the thin and thick disks of the Galactica galaxy when it is at $z=0.3$ and $z=0.0$ and confirmed that the scale height and luminosity ratios between the two disks are similar at the two epochs.

\subsection{Spatial decomposition of thin and thick disks}
\label{sec:decomposition}
A first step towards investigating how distinct the two disk components are is to separate them and compare their properties, e.g., the contribution of stellar particles formed ex situ, stellar age, metallicity, and kinematics. 
In many observational studies, thin and thick disk stars have been separated based on the cut in the chemical plane of [Fe/H] and [$\alpha$/Fe] \citep[e.g.,][]{Lee2011FormationSample, Anders2014ChemodynamicsData, Bensby2014ExploringNeighbourhood, Duong2018TheNeighbourhood, Mackereth2019DynamicalitGaia}. 
Theoretically, some studies also performed decomposition based on the kinematic properties of stellar particles in the simulated galaxies \citep[e.g.,][]{Abadi2003SimulationsDisks, Scannapieco2009TheUniverse, Obreja2018IntroducingGalaxy, Obreja2019NIHAOHaloes}. 
However, it is difficult to spatially separate the thin and thick disk stars, because their distribution overlaps in the disk region; for example, in our simulated galaxy, the thick disk component contributes to roughly 30\% of the stellar mass near the galactic midplane (Figure~\ref{fig:projected_map} d).

In this study, we attempt to separate the thin and thick disk stars based on the $\sl r$-band vertical profiles measured at $R=2\,R_{\rm d}\pm1\,\rm kpc$. 
First, we measured the ``cross height'' ($z_{\rm cross}$) where the contributions from the thin and thick disks equal.
To avoid high contamination from each component near the transition zone (close to $z_{\rm cross}$), we classify all stellar particles below $0.8\,z_{\rm cross}$ as the thin-disk stars and stellar particles located in the region $1.2\,z_{\rm cross}<\rm |Z|<2\,{\it z}_{\rm thick}$ as the thick-disk stars.
Note that while the upper limit of the height for the thick disk is set to $2\,z_{\rm thick}$ to account for the different thick disk extent of each galaxy, the conclusion does not change even when we set the limit to a fixed value of $\rm 2\,kpc$ as adopted in some previous studies.
We applied this decomposition technique to the {\gal} galaxy at $z=0.0$ and the 18 NH galaxies at $z=0.3$ to separate their thin and thick disks.

Note that when spatially separated, the two disks contain overlapping components from each other. For example, in Figure~\ref{fig:projected_map}, we have shown that even in the galactic midplane, thick disk stars contribute to $\sim30\%$ of mass and $\sim10\%$ of luminosity. This would make the difference in properties between the two disks less dramatic, which will be discussed in the next section. However, the spatially-defined thick disk component would be less affected by these overlaps than the thin disk, as the off-plane regions are highly dominated by thick-disk stars. Also, in Section~\ref{sec:evol}, we track the vertical distribution of each mono-age population of stars, without any spatial separation of thin/thick disk, and show that stars are born with thinner vertical distributions but later get thicker with time due to heating. Therefore, our qualitative conclusion is not heavily affected by the overlapping components.

\subsubsection{Decomposition results: age/metallicity and the origins of stellar particles}
\label{sec: decomposition_age_metal}
\begin{figure}
    \linespread{1.0}\selectfont{}
    \includegraphics[width=\columnwidth]{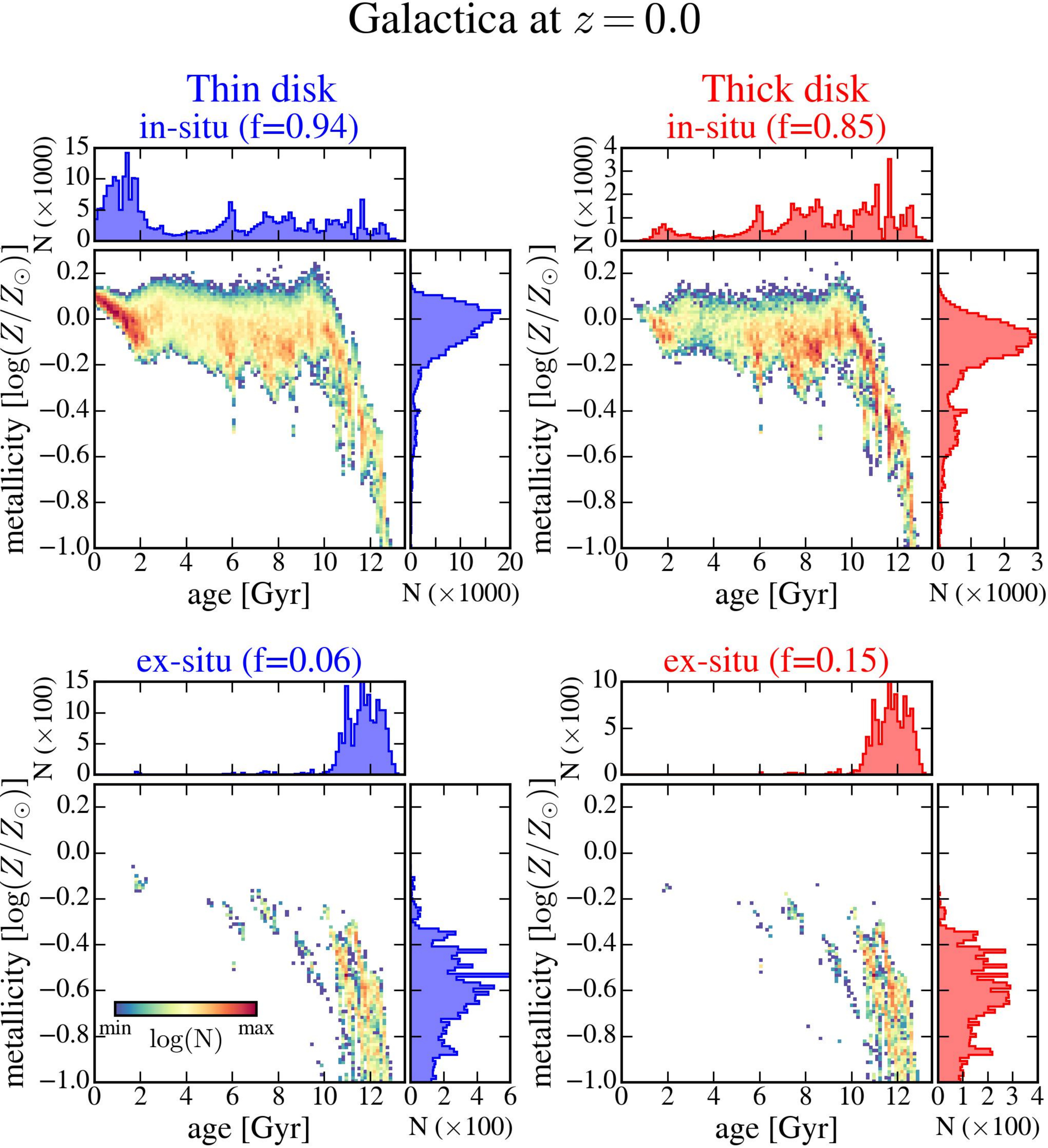}
    \caption{The distribution of the decomposed thin (left, blue histograms) and thick disk stars (right, red histograms) of the {\gal} galaxy} in the plane of age (in $\rm Gyr$) and metallicity (\metallicity). The stellar particles in each component are further divided into in-situ (top panels) and ex-situ stellar particles (bottom panels), and their fraction to each component is shown at the top of each panel. We find that the majority of the stars in the thin ($\sim 95\%$) and thick ($\sim 85\%$) disks are the stellar particles formed in situ. The major difference between the thin and thick disks is the age distribution of the in-situ stars.
    \label{fig:age_metal}
\end{figure}

Figure~\ref{fig:age_metal} shows the age-metallicity distribution of the decomposed thin and thick disk stellar particles in the {\gal} galaxy.
The stellar particles in the thin and thick disks are further divided by their birthplace: in situ or ex situ\footnote{At each snapshot, we tagged stars younger than $50\,\rm Myr$ inside $R_{90}$ of the main progenitors of the galaxy in question as stars formed in situ. Stars that have not been tagged at all until the final epoch are considered as ex-situ formed stars.}. 
The majority ($\sim$95\%) of the thin disk consists of the stars formed in situ, many of which formed recently within the last $\rm 2\,Gyr$ ($\sim$37\%).
On the other hand, $\sim$15\% of the stellar particles in the thick disk are formed ex situ and later accreted, mostly around $z\sim2$ ($\rm 10-12\,Gyr$) during a couple of major merger events.
Even so, the dominant origin of the thick disk is still the stars formed in situ (85\%).
However, the important difference is in the age of the in-situ stars: the age distribution of the in-situ stars of the thick disk is significantly skewed toward larger ages compared to that of the thin disk.

The accretion-origin of the thick disk was investigated by \cite{Abadi2003SimulationsDisks} where they found in their simulation that $\sim$60\% of the stellar particles in their kinematically decomposed thick disk came from the disrupted satellite galaxies.
Indeed, there is one galaxy, FGC227, that has a thick disk with a large fraction of counter-rotating stars which is thought to be originated from accretions \citep[e.g.,][]{Yoachim2008LickGalaxies}.
These accreted stars in the thick disk are predicted to have highly dispersion-dominated kinematics compared to the stars formed in situ \citep[e.g.,][]{Sales2009OrbitalScenarios}.
However, the kinematics of MW thick disk stars seems to disfavor the accretion-origin \citep[e.g.,][]{Lee2011FormationSample, Ruchti2011ObservationalOrigins, Ruchti2014TheDisc}, and, furthermore, recent observational studies on the kinematics of the thick disks in local galaxies have also provided evidence supporting the scenario in which most of the thick stars are formed in situ \citep[e.g.,][]{Comeron2019, Pinna2019TheAccretion}.
Many other simulations have also suggested much lower fractions of ex-situ stars in the thick disks, typically less than a few percent \citep[e.g.,][]{Brook2012ThinGalaxy, Obreja2019NIHAOHaloes}.
For the bulk of our {\gal} and NH galaxies, the accretion of the stars from disrupted satellite galaxies appears to be insignificant.
The median ex-situ fractions of the thin and thick disks in the 18 NH galaxies are $f_{\rm ex\,situ} = 0.06^{+0.05}_{-0.03}$ (thin disks) and $0.11^{+0.16}_{-0.06}$ (thick disks); only two out of 18 galaxies have $f_{\rm ex\,situ} > 0.3$, one of which has $f_{\rm ex\,situ}>0.5$.

The stellar particles in the separated thin and thick disks show differences in age and metallicity. 
The median age of the thin and thick disks in the {\gal} galaxy at $z=0$ is $\rm 5.4^{+5.2}_{-4.4}\,Gyr$ and $\rm 9.6^{+2.2}_{-3.6}\,Gyr$, respectively, with errors derived from the 16th to 84th percentile.
The thin and thick disk components have median metallicity of $\rm \big\langle log(\it {Z/Z_{\odot}} ) \big\rangle \rm =-0.05^{+0.09}_{-0.21}$ and $\rm \big\langle log( \it{ Z/Z_{\odot}} )\big\rangle \rm=-0.16^{+0.12}_{-0.44}$, where $Z_{\odot}=0.0134$ \citep{Asplund2009TheSun}. 
The Table~\ref{table:properties} summarizes the differences in age and metallicity between the thin and thick disks in the {\gal} galaxy at $z=0.0$ (the first column) and also those for the MW galaxy (the thrid column).

\begin{table*}[ht]
\centering
\caption{Properties of the thin and thick disks in the {\gal} disk galaxy at $z=0.0$ (the first column), 18 NH galaxies at $z=0.3$ (the second column), and the Milky Way (the third column).} 
\label{table:properties}
\begin{threeparttable}
\begin{adjustwidth}{-1cm}{}
       \makebox[1 \textwidth][c]{   
       \resizebox{1 \linewidth}{!}{

\begin{tabular}{p{0.27\textwidth}>{\centering\arraybackslash}p{0.16\textwidth}>{\centering\arraybackslash}p{0.2\textwidth}>{\centering\arraybackslash}p{0.17\textwidth}}
\hline
  & {\gal} ($z=0.0$) 
  & $<$18 NH galaxies$>$ ($z=0.3$, $t_{\rm lb}\tnote{*}\sim3.5\,\rm Gyr$)
  & MW \\ 
\hline
\hline
$M_{\rm stellar}$ [$\times10^{10}\,\msun$]
& $2.8$ 
& $1.7$ 
& $5.0-7.0$ \tnote{a} \\ \hline

Thin disk age [Gyr]
& $5.4^{+5.3}_{-4.4}$ 
& $4.0^{+1.5}_{-1.8}$ 
& $\rm <8$ \tnote{b,c}\\

Thick disk age [Gyr]
& $9.7^{+2.2}_{-3.7}$ 
& $5.6^{+1.0}_{-1.9}$ 
& $>8-10$ \tnote{b,c} \\

$\Delta$ age (median) [Gyr]
& $4.3$ 
& $1.6$ 
& - \\\hline 

Thin disk age ($\sl r$-band) [Gyr]
& $2.2$ 
& $1.7$ 
& - \\

Thick disk age ($\sl r$-band) [Gyr]
& $7.8$ 
& $4.7$ 
& - \\

$\Delta$ age ($\sl r$-band) [Gyr] 
& $5.7$ 
& $3.0$ 
& - \\\hline

Thin disk $\rm log(\it Z/Z_{\rm \odot})$ 
& $-0.05$ 
& $-0.06$ 
& $\rm [Fe/H]\sim-0.2$ \tnote{d} \\

Thick disk $\rm log(\it Z/Z_{\rm \odot})$ 
& $-0.17$ 
& $-0.22$ 
& $\rm [Fe/H]\sim-0.6$ \tnote{d} \\

$\Delta \rm log(\it Z/Z_{\rm \odot})$ (median) [dex]
& $0.12$ 
& $0.16$ 
& $0.4$ \\\hline

$\Delta\,\rm log(\it Z/Z_{\rm \odot})$ ($\sl r$-band) [dex]
& $0.20$ 
& $0.15$ 
& - \\\hline

Thin disk $V_{\rm rot}$ [$\rm km/s$]
& $172^{+54}_{-79}$ 
& $109^{+66}_{-21}$ 
& $220$\\

Thick disk $V_{\rm rot}$ [$\rm km/s$]
& $129^{+73}_{-115}$ 
& $89^{+54}_{-23}$ 
& $180$ \\

$\Delta\,V_{\rm rot}$ (median) [$\rm km/s$] 
& $43$ 
& $20$ 
& $20-70$ \tnote{d,e} \\\hline

Thin disk $f_{\rm ex\,situ}$ 
& $0.06$ 
& $0.06^{+0.05}_{-0.03}$ 
& - \\

Thick disk $f_{\rm ex\,situ}$ 
& 0.15
& $0.11^{+0.16}_{-0.06}$ 
& - \\
\hline
\end{tabular}}}
\begin{tablenotes}
\small
  \item Note. -- For the {\gal} galaxy, all the given values are the median/$\sl r$-band weighted average values of the stellar particles with errors showing 16th to 84th percentiles. The values of each NH galaxy is measured in the same way, and the median values of the 18 NH galaxies are presented in the second column, with errors showing from 16th to 84th percentiles among the sample galaxies.
  \item[*] $t_{\rm lb}$: Look back time from $z=0.0$ is $\sim3.5\,\rm Gyr$
  \item[a] \cite{McMillan2011} 
  \item[b] \cite{Bensby2014ExploringNeighbourhood}
  \item[c] \cite{Haywood2013TheDisk}
  \item[d] \cite{Lee2011FormationSample}
  \item[e] \cite{Anders2014ChemodynamicsData} 
\end{tablenotes}
\end{adjustwidth}
\end{threeparttable}
\end{table*}

The ages of the thin and thick disks (and the difference between the two components) are consistent with those of MW's \citep[e.g.,][]{Haywood2013TheDisk}.
However, the metallicity difference between the two components of the {\gal} galaxy, $\Delta \rm [Fe/H] \approx -0.12\,\rm dex$\footnote{We here assume that Fe scales with $Z$ and the difference in the hydrogen abundance is negligible.}, seems significantly smaller than that found in the MW. For example, when the thin and thick disks of the MW are decomposed based on the chemical compositions, the metallicity difference has been found to be $\Delta \rm [Fe/H]\sim-0.4\,\rm dex$ \citep[e.g.,][]{Lee2011FormationSample}. 
One thing we need to consider is that the stellar mass of the {\gal} galaxy is only half that of the MW, so, in general, the galaxy might not be as enriched as the MW \citep[e.g.,][]{Faber1973, Gallazzi2006}.
The other issue is the merger history of the {\gal} galaxy; it has a significant major merger with a mass ratio of $0.24$ at $z\sim2.2$, followed by rapid mass growth and metal enrichment. 
This is why the thick disk component, even considering only the stellar particles formed in situ (upper right panel in Figure~\ref{fig:age_metal}), has a bimodal distribution of metallicity with a mild peak at $\rm log(\it{Z/Z_{\odot}})\sim \rm-0.5$ and a much stronger peak at $\rm log(\it{Z/Z_{\odot}})\sim \rm -0.1$.
On the other hand, the enhanced star formation between $\rm 0-2\,Gyr$, most of which contributes to the growth of the thin disk, as shown in the upper-left panel (see also Panel (a) of Figure~\ref{fig:hz_evolution}) is driven by the interaction with two satellite galaxies, consistently with observations \citep[see also][]{Ruiz-Lara2020TheHistory}, from which the metal-poor gas is accreted as well. 
This might have diluted the interstellar medium in the galaxy, thus leading to a small difference in metallicity between the younger stellar particles (with age $\rm \sim0-2\,Gyr$) and the older stellar particles (with age $\rm \sim8-10\,Gyr$).

Also, when comparing the values between the Galactica galaxy and the MW, it is important to note that different definitions of thin/thick disk are adopted, which might affect the differences between the two disks; 
When the thin disk is defined chemically, as in many of the MW studies \citep[e.g.,][]{Lee2011FormationSample,Bensby2014ExploringNeighbourhood,Anders2014ChemodynamicsData}, ``young'' thin disk stars can be well separated, as the alpha abundances broadly trace the age of the populations. 
Since younger stars tend to be more metal-rich and have higher rotational velocity, the difference in properties between the thin and thick disks would be smaller when defined based on spatial cut than when the two disks are chemically defined. 
Indeed, when we separate the two disks based on a simple ``age'' cut of 8 Gyr, for example, the differences in properties obviously increase to $\rm \Delta age\sim7.1\rm\,Gyr$, $\Delta \log(Z/Z_{\odot})\sim0.32\rm\,dex$, and $\Delta V_{\rm rot}\sim81\rm\,km/s$. 
However, this age cut is based on the assumptions that we know a priori that the thick disk stars formed earlier consisting of older stars, regardless of their origins (e.g., in situ or ex situ). 
On the other hand, when the thick disk is defined spatially, as in this study, we focus on the ``thickly-distributed'' component primarily occupying higher altitudes from the galactic midplane, with the same motivation that the concept of a thick disk was first proposed to explain the vertical distribution.

We have performed the same analysis on the 18 NH galaxies at $z=0.3$.
The mass-weighted (or $\sl r$-band-weighted) age and metallicity difference in each galaxy is calculated from the median values (weighted average) of the thin and thick disk stellar particles.
The median values of the properties of the two disks in the 18 NH samples at $z=0.3$ ($t_{\rm univ}\sim\,\rm 10\,Gyr$) are also given in Table~\ref{table:properties}.

Similarly, the thick disk components in the NH galaxies are older and metal-poorer than the thin disk components, which is qualitatively consistent with observations \citep[e.g.,][]{Yoachim2008LickGalaxies,Lee2011FormationSample, Pinna2019TheEnvironment, Kasparova2020ANGC7572}, while the metallicity differences remain small compared to observations.
The difference, in age and metallicity, would likely be larger when measured at $z=0$ in typical star-forming galaxies, because continued star formation is likely to increase the mean metallicity of young stars in the thin disk. 
However, the exact amount will depend on the detailed formation history of each galaxy.

\subsubsection{Decomposition results: kinematics}
\label{sec: decomposition_kinematics}
The thick disk of the MW defined by chemical abundances is thought to have different kinematic properties from the thin disk with a slower rotational velocity by 40-50\,\rm km/s \citep[see][]{Lee2011FormationSample, Anders2014ChemodynamicsData}.
Also, similar lags in velocity have been found in simulations where the thick disk stars are selected based on age \citep[see][]{Brook2012ThinGalaxy, Minchev2013ChemodynamicalVicinity}.
While considering this as a significant difference, some studies have identified the thin and thick disks according to the kinematic properties, for example, using the so-called Toomre diagram \citep[see][]{Bensby2003ElementalStars}.
Here we explore the velocity difference between the two components, in our case, separated by the spatial definition, and analyze the cause of the difference by tracking the origin of stellar particles belonging to each component.

\begin{figure*}
    \linespread{1.0}\selectfont{}
    \includegraphics[width=\textwidth]{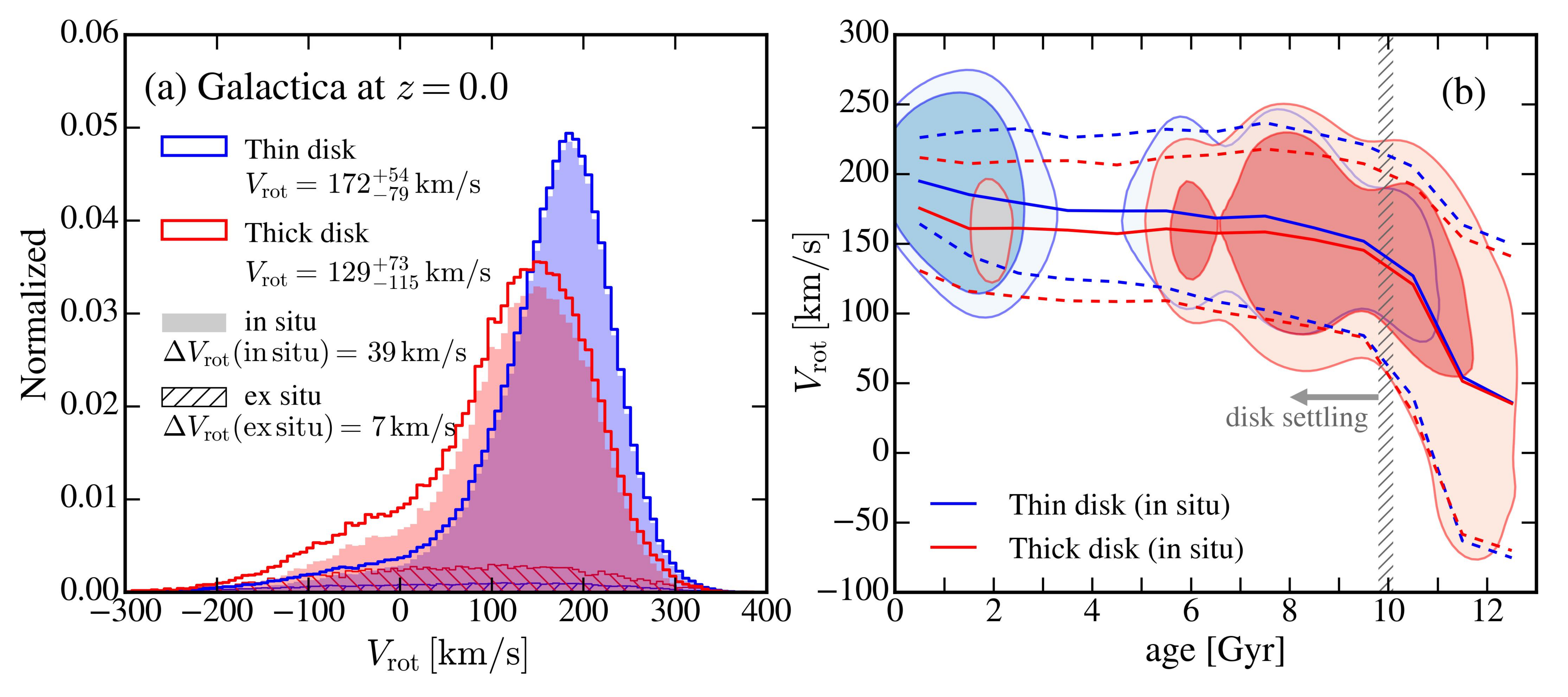}
    \caption{(a) The distribution of the rotational velocity ($V_{\rm rot}$) of the thin (blue) and thick (red) disk stellar particles in the {\gal} galaxy at $z=0.0$, normalized by the total number of stellar particles in each component. The stellar particles in each component are further divided into stars formed in situ (solid histogram) and ex situ (hatched histogram). The difference in rotational velocity between the two components ($\Delta V_{\rm rot}=43\,\rm km/s$) can be largely explained by the velocity difference between the in situ stellar particles in each component ($\Delta V_{\rm rot}\,(\rm in\,situ)=39\,\rm km/s$). (b) The rotational velocity of the thin (blue line) and thick (red line) disk stellar particles formed in situ as a function of stellar ages. The age distribution of the in situ thin and thick disk stars are shown as blue and red contours showing $0.5\,\sigma$ and $1\,\sigma$. The vertical hatched line is the disk settling epoch ($z_{\rm disk\,settling}$) when the galaxy starts to develop its disk structure. 
    We find that the rotational velocity of the stars formed in situ increases with decreasing ages, and it is the different age distribution between the thin and thick disks that contributes the most to the difference in $V_{\rm rot}$ between the two components.}
    \label{fig:vrot}
\end{figure*}

Figure~\ref{fig:vrot} (a) shows the rotational velocity distribution of the thin and thick disk stellar particles in the {\gal} galaxy, normalized by the total number of stellar particles in each component. 
The rotational velocity is measured as the tangential velocity, $V_{\phi}$, of the stellar particles in the cylindrical coordinate of the galaxy where the z-axis is defined as the rotational axis of the galaxy.
The rotational velocity of the thin disk component is $\big\langle V_{\rm rot,\, thin}\big\rangle =172^{+54}_{-79}\,\rm km/s$ (median and the error showing from 16th to 84th percentile), while the thick disk component exhibits a velocity lag of $\rm \sim43\,km/s$ ($\big\langle V_{\rm rot,\, thick}\big\rangle=129^{+73}_{-115} \,\rm km/s$). 
We also further divide the stellar particles in each component into stellar particles formed in situ and ex situ and measure their rotational velocities. 
We find that the velocity lag of the thick disk is caused mostly by the in-situ stellar particles ($\Delta V_{\rm rot}\,(\rm in\,situ)=39\,\rm km/s$) rather than the accreted stellar particles \citep[see][]{Qu2011MinorGalaxies}.

Panel (b) shows the median rotational velocity ($V_{\rm rot}$) of the ``in situ'' thin (blue solid line) and thick (red solid line) disk stars as a function of their ages. 
The blue/red dashed lines show the 16th and 84th percentiles of the $V_{\rm rot}$ at a given stellar age of the thin/thick disk stars. 
The distributions of the in-situ thin and thick disk stars in the plane of age and $V_{\rm rot}$ are shown as blue and red contours ($0.5\,\sigma$ and $1\,\sigma$). 
The vertical hatched line represents the disk settling epoch ($z_{\rm disk\,settling}$) when the galactic disk structure starts to develop. 
To quantitatively define this epoch, we used $V/\sigma$ of the cold gas ($n_{\rm H} > \rm 10\,cm^{-3}$ and $T<2\times10^4\,\rm K$) inside $R_{90}$ of the galaxy. 
The $V/\sigma$ is measured as the mass-weighted mean tangential velocity ($V=<V_{\phi}>$) of the cold gas cells in the cylindrical coordinates, where the z-axis is defined as the direction of the net angular momentum of the cold cells, divided by mass-weighted 1D-velocity dispersion: 
$\sigma = \sqrt{(\sigma_{r}^2 + \sigma_{\phi}^2 + \sigma_{z}^2)/3}$.
The $z_{\rm disk\,settling}$ is defined as the time when the mean cold gas $V/\sigma$ for the past $150\,\rm Myr$ ($\sim 10\,\rm snapshots$) first reaches higher than 3, which is used as a criterion for settled disks in some studies \citep[e.g.,][]{Kassin2012TheFormation} (See also Dubois et al. in prep).

The rotational velocity of the stars formed in situ increases with decreasing age. 
Especially, the stellar particles formed in situ before the disk settling epoch (right side of the vertical hatched line) have much lower rotational velocity, compared to the young stars formed recently, because the galaxy was much smaller at higher redshifts. 
It could be also attributed to the fact that many of the stars formed before disk settling were formed with random-dominated kinematics during chaotic mergers which might have prevented the galaxy from settling its disk. 
What contributes the most to the difference in $V_{\rm rot}$ between the thin and thick disks is the difference in age distribution between the stars in the two components: while many of the thin disk stars are young stars with high $V_{\rm rot}$, appoximately $40\%$ of the in-situ thick disk stars are formed before the disk settling, contributing to the lower end of $V_{\rm rot}$ distribution of the thick disk.
Also, most of the thick disk stars formed after the disk settling are old stars with ages $\rm 6-10\,Gyr$. 
One noticeable feature is that for the stars formed after the disk settling ($<10\,\rm Gyr$), thick disk stars have lower $V_{\rm rot}$ than the thin disk stars with the same age, which may have resulted from disk heating.
The difference in $V_{\rm rot}$ for the stars formed in situ after the disk settling is $\Delta V_{\rm rot}$ (in-situ, after disk settling) $=25\,\rm km/s$.

\begin{figure*}
    \linespread{1.0}\selectfont{}
    \includegraphics[width=\textwidth]{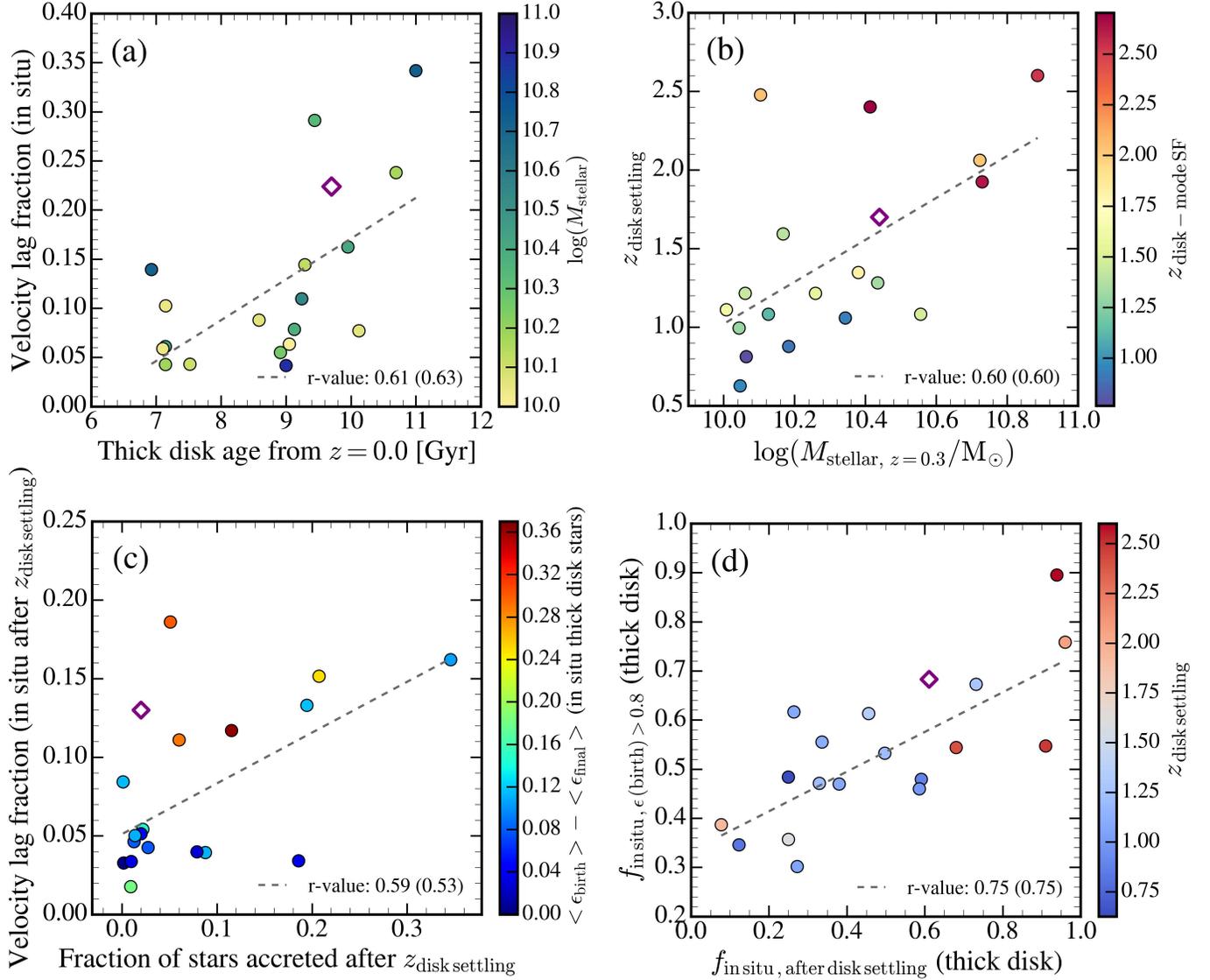}
    \caption{(a) The velocity lag fraction for the ${\rm in\,situ}$ stellar particles in the thin and thick disks in the 18 NH galaxies (at $z=0.3$) as a function of the age of the thick disks. The age of the thick disks is taken as the median age of the stellar particles in the thick disk component plus $t_{\rm look\,back}=3.45\,\rm Gyr$ to obtain their ages from $z=0.0$. The galaxies are color-coded according to their stellar mass. 
    (b) The disk settling epoch ($z_{\rm disk\,settling}$, defined as the time when the mean cold gas $V/\sigma$ for the past $150\,\rm Myr$ first reaches higher than 3, see the text for details) as a function of stellar mass of the galaxies at $z=0.3$. The galaxies are color-coded by the epoch when a galaxy starts to pre-dominantly form disk stars ($z_{\rm disk-mode\,SF}$, see the text). 
    (c) The velocity lag fraction for the stellar particles formed in situ after $z_{\rm disk\,settling}$ as a function of the fraction of stars accreted after $z_{\rm disk\,settling}$ in the total number of stars in a galaxy. The color represents the change in orbital circularity of the thick disk stars formed in situ between at their birth and at $z=0.3$ (the final epoch). 
    (d) The fraction of stars in the thick disk that are formed in situ with $\epsilon_{\rm birth}$ (orbital circularity at birth) greater than 0.8, as a function of the fraction of stars in the thick disk that are formed in situ after $z_{\rm disk\,settling}$. The color indicates the $z_{\rm disk\,settling}$. 
    All in the four panels, the values of the {\gal} galaxy at $z=0.0$ are shown in as open purple diamonds. The gray dashed line in each panel is the linear fit to the data of the 18 NH galaxies with Pearson correlation coefficient (r value) given in the lower right. The value in the parentheses in each panel is the correlation coefficient obtained when the value of the {\gal} galaxy is also included. 
    We find that different age distributions of the ``in-situ'' stars in the thin and thick disks contribute greatly to the difference in rotational velocity. 
    Thick disks rotate slower than thin disks because some of them were formed with random-dominated kinematics before the disk settling, while others formed after disk settling were also slowed down due to heating.}
    \label{fig:vlag_fraction}
\end{figure*}

We have also measured the velocity lag for the 18 NH galaxies. 
We define the ``velocity lag fraction'' as the velocity lag normalized by the circular velocity ($V_{\rm circ}$) of each galaxy to compare the degree of kinematic differences in galaxies with different masses:
\begin{equation}
    \rm Velocity\,lag\,fraction = (\textit{V}_{rot,\,\rm thin} - \textit{V}_{rot,\,\rm thick})/\textit{V}_{circ}
\end{equation}
where $V_{\rm circ}$ is the circular velocity measured at $R_{90}$ of the galaxy following $V_{\rm circ}=\sqrt{GM(<r)/r}$.
All the NH galaxies have a positive velocity lag fraction, which can be understood accordingly to the vertical gradient of the mean rotational velocity found in several observations \citep[e.g.,][]{Ivezic2008TheMetallicity, Carollo2010StructureDR7} and simulations \citep[e.g.,][]{DiMatteo2011TheNeighborhood}.
However, the velocity lag varies across galaxies even after mass (circular velocity) normalization. 
The median value for the NH galaxies at $z=0.3$ is 0.08 with a range from 0.06 (16th) to 0.22 (84th percentile).

There are several elements to consider to understand the velocity lag fraction; (i) first, the amount of stars formed ex situ contributes to the lower end of the distribution of $V_{\rm rot}$, as the accreted stars are thought to have dispersion-dominated kinematics with highly eccentric orbits \citep[e.g.,][]{Sales2009OrbitalScenarios}. 
However, as the amount of ex-situ stars is low for most of the galaxies in our simulations, their contribution to the total kinematics would also not be significant. (ii) Second, stars formed in situ before disk settling would have low rotational velocity, as seen in Figure~\ref{fig:vrot} (b). (iii) Finally, stars formed in situ after disk settling could have high $V_{\rm rot}$ at their birth, but their rotational speed could decrease with time due to heating.

First, we examine how the different age distributions of the thin and thick disks (or how old the thick disks are) contribute to the velocity lag of the thick disks. 
Figure~\ref{fig:vlag_fraction} (a) shows the velocity lag fraction using only ``${\it in\,situ}$'' stellar particles as a function of the age of the thick disk. 
The age of the thick disk was measured from $z=0.0$ by adding $3.45\,\rm Gyr$ (look back time of $z=0.3$) to the median age of the stellar particles in the thick disks.  
Indeed, there seems a mild correlation between the velocity lag fraction for the in-stiu formed stars and the age of the thick disk with a correlation coefficient of 0.6. 
The color in Panel (a) represents the stellar mass of the NH galaxies at $z=0.3$, which does not seem to have much to do with the velocity lag fractions or the age of the thick disks.

As seen in Figure~\ref{fig:vrot}, stars formed before the disk settling have lower $V_{\rm rot}$, probably because they were formed during chaotic mergers with less ordered motions. 
To find out when the galaxies start pre-dominantly forming stars in the disk with ordered motions, we measure when the disk structures begin to appear in the NH galaxies (the disk settling epoch, $z_{\rm disk\,settling}$).
As described above, we used a $V/\sigma$ criterion to identify the $z_{\rm disk\,settling}$, and Panel (b) shows the $z_{\rm disk\,settling}$ of the NH galaxies as a function of their stellar mass at $z=0.3$. 
We find the trend that more massive galaxies tend to settle their disks earlier than less massive galaxies, which is consistent with the trend found in observations \citep[e.g.,][]{Kassin2012TheFormation}.
The detailed analysis on disk settling will be addressed in a separate paper.
As a sanity check, we also measure the epoch when the galaxies start to form disk stars predominantly. 
We define the ``disk-mode'' star formation as more than 80\% of the newly-formed young stars ($<100\,\rm Myr$) are disk stars (defined as the stars with $\epsilon>0.5$) and measure the first moment when this disk-mode SF lasts for $150\,\rm Myr$, $z_{\rm disk-mode\,SF}$. 
This $z_{\rm disk-mode\,SF}$ is represented as the color of each marker, and it seems that this agrees well with $z_{\rm disk\,settling}$ defined using gas $V/\sigma$.

One of the mechanisms behind the velocity lag fraction for the stars formed in situ after $z_{\rm disk\,settling}$ would be disk heating. 
Several sources of heating have been suggested including mergers \citep[e.g., ][]{Quinn1993HeatingMergers} as an external source and spiral arms \citep[e.g.,][]{SellwoodJ.A.1984SpiralFormation} and giant molecular clouds \citep[e.g,][]{Spitzer1951TheVelocities} as internal sources. 
In Panel (c), we plotted the velocity lag fraction for the stars formed in situ after $z_{\rm disk\,settling}$ as a function of the fraction of the stars accreted after $z_{\rm disk\,settling}$ in the total number of stars in a galaxy.  
The color in each marker shows the change in orbital circularity of the in-situ thick disk stars between at their birth and at the final epoch ($z=0.3$), $\big \langle \epsilon_{\rm birth} \big \rangle - \big \langle \epsilon_{\rm final} \big \rangle$. 
Broadly, thick disks in the NH galaxies with high velocity lag fraction tend to have much less aligned orbits than at birth.
We also find that there is a slight hint of correlation (with a correlation coefficient of 0.59) between the ex-situ star fraction since disk settling and the velocity lag fraction measured for the stars formed in situ after $z_{\rm disk\,settling}$.
This suggests that mergers contribute to disk heating as an external heating source. 
However, this does not preclude the possibility of heating caused by secular evolution, as trend shown in Panel (c) is not strong, and when mergers or interactions occur, they often trigger star formation, which contributes to promoting heating by internal sources as well.

Finally, in Panel (d), we measured how many of the thick disk stars formed in situ are formed after the disk settling, $f_{\rm in\,situ,\,after\,disk\,settling}$, and compare it with the fraction of thick disk stars formed in situ with orbital circularity higher than 0.8, $f_{\rm in\,situ,\,\epsilon_{birth}>0.8}$.
We find that there is a strong correlation between them.
This means that most of the thick disks stars formed after disk settling are born as kinematically thin-disk stars ($\epsilon_{\rm birth}>0.8$).
In the NH galaxies, a significant fraction of the in-situ thick disk stars are born after the disk settling (mean $\big \langle f_{\rm in\,situ,\,after\,disk\,settling} \big \rangle=0.48$) and as kinematically thin-disk stars (mean $\big \langle f_{\rm in\,situ,\,\epsilon_{birth}>0.8}\big \rangle = 0.53$). 
The color inside each marker indicates the $z_{\rm disk\,settling}$ of each galaxy; in galaxies that developed their disks earlier, a larger fraction of the stars in the thick disk was formed as kinematically thin-disk stars.
It should be, however, noted that the fraction of the in-situ formed stars after disk settling that comprises the thick disk (in Panel d) varies from 10\% to 95\% depending on the details in the star formation and merger accretion history. 
Therefore, it is important to secure a good number of sample galaxies to derive a representative picture of the disk formation.

\subsubsection{Decomposition results: properties at birth}
\label{sec: decomposition_birthprops}
\begin{figure}
    \linespread{1.0}\selectfont{}
    \includegraphics[width=\columnwidth]{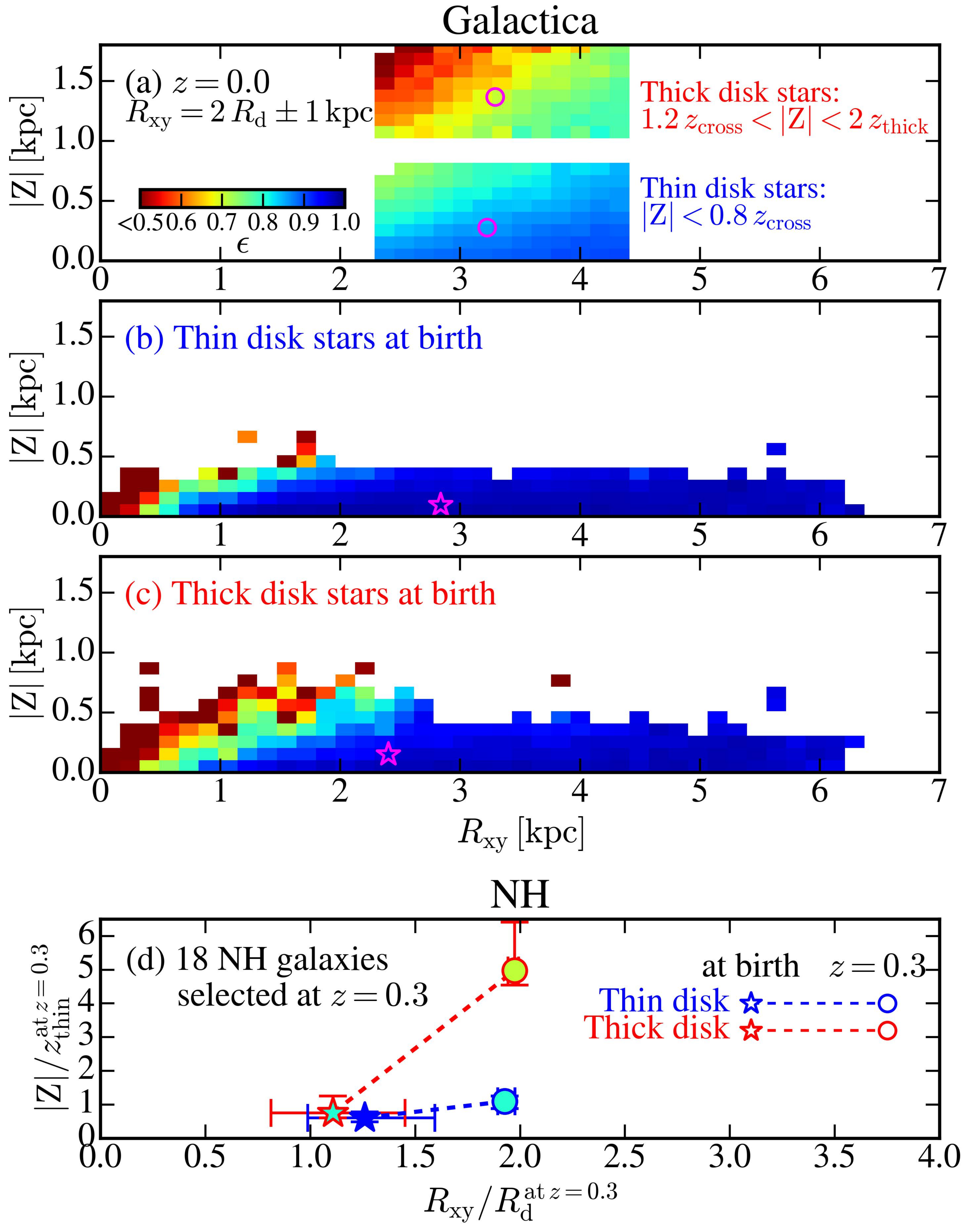}
    \caption{The comparison in the position of ``in-situ'' formed stellar particles in the thin and thick disk at the final epoch and at their birth. (a) The final position (at $z=0.0$) of the stellar particles in the thin and thick disks of the {\gal} galaxy, selected by the definition ($R_{\rm xy}=2\,R_{\rm d}\pm1\,\rm kpc$, thin disk: $\rm |Z|<0.8\,{\it z}_{\rm cross}$, thick disk: $1.2\,z_{\rm cross}<\rm |Z|<2\,{\it z}_{\rm thick}$). The position of the stellar particles in the thin/thick disk components at the time of birth is shown in Panels (b) and (c). Only the pixelated regions containing more than 0.1\% of the total number of stellar particles in each component are displayed. Each pixel is color-coded by the mean orbital circularity ($\epsilon$) of the stellar particles in each region. Panel (d) shows the median changes in the positions of the thin and thick disks for the 18 NH galaxies at $z=0.3$. The blue/red circles present the final ($z=0.3$) positions of thin/thick disks, while the blue/red stars indicate their birthplaces. To consider different sizes/masses of the NH sample galaxies, we normalized the projected radial distance of each galaxy and by the scale length of its disk and the vertical distance by the scale height of the thin disk. We find that a significant fraction of the thick disk stars were spatially and kinematically thinner at birth.}
    \label{fig:compare_birth_positions}
\end{figure}

In the previous section, we quantified the difference in rotation speed between the two disks extracted from the two non-overlapping disk regions of the galaxies. 
In the light of exploring whether they are distinct components put in place by different mechanisms, here we explore the properties of the stellar particles in the two components at birth: their kinematic properties at birth and their birthplaces. 
For the {\gal} galaxy, Figure~\ref{fig:compare_birth_positions} (a) shows the final ($z=0$) position of the stellar particles in the two components selected by the definition ($R_{\rm xy}=2\,R_{\rm d}\pm1\,\rm kpc$, thin disk: $\rm |Z|<0.8\,{\it z}_{\rm cross}$, thick disk: $1.2\,z_{\rm cross}<\rm |Z|<2\,{\it z}_{\rm thick}$). 
The magenta circles show the median positions (in the plane of $R_{\rm xy}$ and $\rm |Z|$) of the stellar particles in the two selected components.
The color represents the median orbital circularity \citep{Abadi2003SimulationsDisks} of the stellar particles residing in each region.

The birth positions of the stellar particles belonging to the thin and thick disks (separated at $z=0.0$) are shown in Panels (b) and (c). 
Note that only the stellar particles formed ``in-situ'' are taken into account when tracking their birthplaces.
Only regions containing more than 0.1\% of the total number of stellar particles in each component are displayed.  
The magenta star in each panel points to the median birth position of each component.
The stellar particles in both components formed over a wide range of radial distances, and most of them were formed in the inner regions ($R_{\rm xy}=2$-3\,\rm kpc, magenta star) and have migrated outward to the final radial distance ($R_{\rm xy}=3.4\,\rm kpc$).   
Also, the majority of them formed close to the galactic midplane ($\rm |Z|<0.5\,\rm kpc$) and they were kinematically much thinner at birth with orbital circularity ($\epsilon$) close to 1 (represented as bluer color).

Panel (d) of Figure~\ref{fig:compare_birth_positions} shows the median changes in the positions across the thin and thick disks in the 18 NH galaxies selected at $z=0.3$. 
The blue/red circles are the final ($z=0.3$) positions of the stellar particles in the thin/thick disks, and the blue/red stars show where these same stellar particles were formed. 
To account for the different sizes of the sample galaxies, we normalized their radial distances by the scale lengths of the disk at the final epoch (i.e., $R_{\rm xy}/R_{\rm d}$) and the vertical distances by the scale heights of the thin disk measured at $z=0.3$ (i.e., $|\rm Z|/{\it z}_{\rm thin}$).
The error bars in the x and y-axes also indicate the 16th to 84th percentiles of the radial distances and the vertical distances of the 18 NH galaxies. We find a consistent result for both {\gal} and NH galaxies. 
Both thin and thick disk stars were born close to the galactic center and the midplane and migrated outward and upward (in this diagram). 
As the bulk of the stellar particles in the two components moves outward and upward, their orbits become less aligned (see the color change inside the markers between the time of birth and $z=0.3$). 
\textit{This suggests that the spatially thicker disk in most NH galaxies was not formed by a distinct mechanism but a result of the time evolution of the stars that were born in a thinner distribution with nearly circular orbits.}

\subsubsection{Decomposition results: Summary}
In Section~\ref{sec:decomposition}, we focused on the difference in various properties between the two spatially separated components. 
We found that both thin and thick disks are dominated by in-situ formed stars, although thick disks have higher fraction of stellar particles formed ex situ.
Thus, the accretion scenario for the thick disk formation is less favored by our simulations. 
We also showed that thin disks are younger, metal-richer, and rotating faster than thick disks, which is qualitatively consistent with observations. 
Especially, the important difference between the thin and thick disks is the age distribution of the in-situ stars; while many of the thin-disk stars are formed recently with rotation-dominated kinematics, thick-disk stars are much older and rotating slower. This is because approximately half of the in-situ thick-disk stars were formed with dispersion-dominated kinematics before galaxies develop their disks, possibly during the mergers \citep[e.g.,][]{Brook2004TheUniverse}, and the other half were formed on the disk but slowed down by heating \citep[e.g.,][]{Quinn1993HeatingMergers}.
However, the exact contribution of each process varies greatly across the galaxies and depends on the detailed star formation and merger history.
In addition, we explored the distribution and kinematic properties of the stellar particles in the two disks at birth. 
We found that a significant part of the thick disk was spatially and kinematically thin at birth.
All in all, this suggests that the two disks are not entirely {\em distinct} components in terms of formation process but they are rather snapshots in time and space of a {\em continuous} evolution of a galactic disk. 
In the next section, we will investigate the evolution of the disks and discuss their origins.


\section{Evolution of the vertical structure of disks} 
\label{sec:evol}
Many theoretical studies have pointed out that galactic disks grow in an ``inside-out'' and ``upside-down'' fashion \citep[e.g.,][]{Stinson2013MaGICCWay,Bird2013InsidePopulations,Martig2014DissectingPopulations,Minchev2015OnDisks,Bird2020}: A galaxy forms a much thinner (in vertical) and more extended (in radial) disk with time, as gas settles into a galaxy. 
In this context, to further understand the origin of the two disks derived from fitting the vertical profiles, this section will focus on how the disk vertical structures evolve with time.
We divide the stellar particles formed in situ into different age groups and see the change in the spatial distribution of each mono-age group.

\subsection{Stellar vertical profile evolution: Galactica}
\label{sec:evol_Galactica}
\begin{figure*}
    \linespread{1.0}\selectfont{}
    \includegraphics[width=\textwidth]{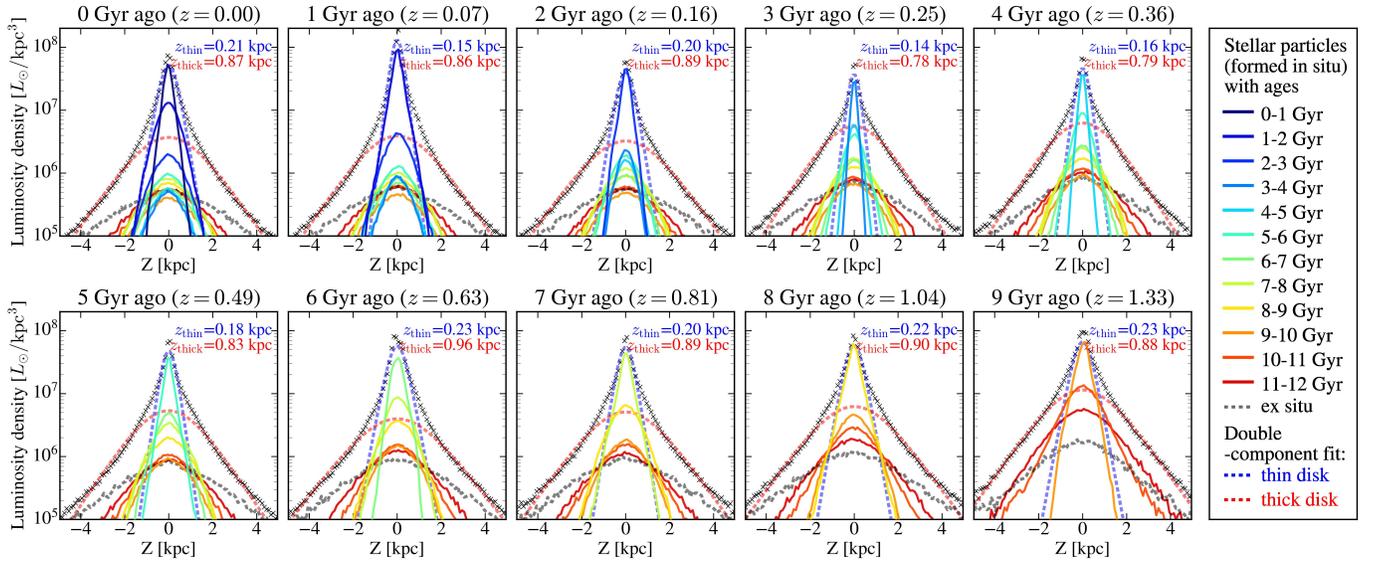}
    \caption{The evolution of the vertical profile of the {\gal} galaxy (measured in a cylindrical region at $2\,R_{\rm d}\pm1\,\rm kpc$) from $z=1.3$ ($\rm 9\,Gyr$ ago) to $z=0.0$ ($\rm 0\,Gyr$ ago).
    The vertical luminosity density is plotted as the black crosses. 
    Each solid line shows the vertical distribution of each mono-age group of stellar particles formed in situ. 
    The gray dotted line is the distribution of stellar particles formed ex situ. 
    The double-component fitting to the vertical profile is represented as dashed blue (thin disk) and red (thick disk) lines. The resulting scale heights of thin and thick disks are displayed on the top of each panel. At all redshifts, the thin disk is dominated by the younger population, while each mono-age population gets thicker and fainter with time, building up the thick disk.
    }
    \label{fig:vertical_profile_evolution}
\end{figure*}

Figure~\ref{fig:vertical_profile_evolution} shows how the vertical distributions (measured in the cylindrical region at $2\,R_{\rm d}\pm1\,\rm kpc$ at each redshift) in $\sl r$-band change over time in the {\gal} galaxy. 
Each panel shows the vertical profile at the corresponding epoch, represented as black crosses. 
The dotted gray lines represent the vertical distributions of stellar particles formed ex situ, which are much thicker compared to the in-situ stellar distributions and do not change much over time.
The distribution of the mono-age populations formed in situ, indicated as solid lines, becomes thicker with time. 
As a result, at $z=0$, younger stellar particles (bluer lines) appear to have shorter scale heights than older stellar particles (redder lines). 
The increasing scale height with stellar age is also found in many previous simulation studies \citep[e.g.,][]{Martig2014DissectingPopulations, Ma2017TheGalaxy, Buck2020NIHAO-UHD:Simulations}, and this trend is also clearly visible among the mono-abundance populations in the MW \citep[e.g.,][]{Bovy2012TheDisk}.

\begin{figure}
    \linespread{1.0}\selectfont{}
    \includegraphics[width=0.8\columnwidth]{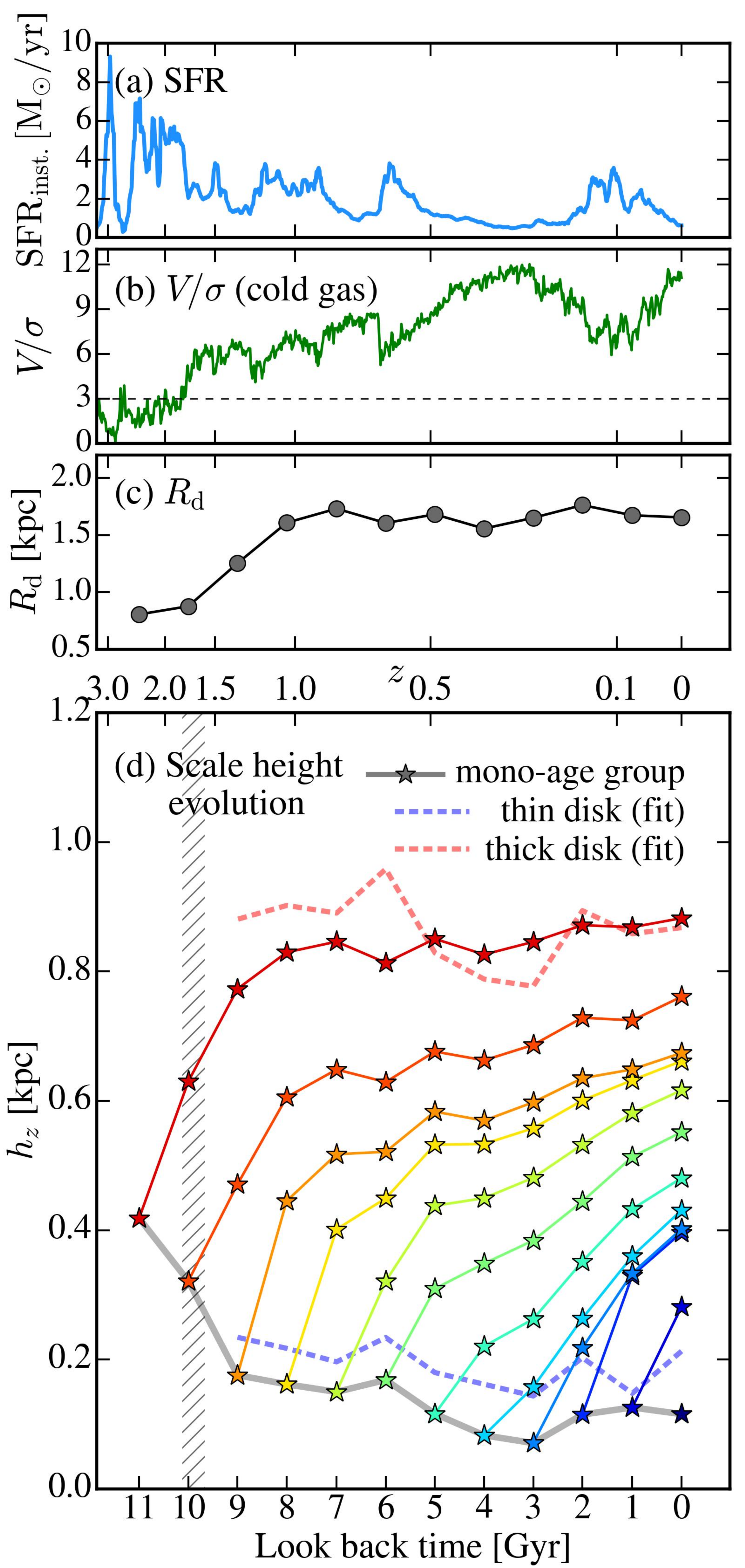}
    \caption{Evolution of the vertical distribution of the disk in the {\gal} galaxy. 
    (a) The instantaneous star formation rate (SFR) as a function of redshift. (b) The evolution of the $V/\sigma$ of the cold gas in the galaxy. 
    (c) The evolution of disk scale length ($R_{\rm d}$) of the galaxy.
    (d) The scale height evolution of mono-age groups of stellar particles indicated as different colors from red to blue with age bin of $\rm1\,Gyr$ (the same color key in Figure~\ref{fig:vertical_profile_evolution}). 
    The vertical distribution is measured at $2\,R_{\rm d}$ of the galaxy at each epoch. 
    The gray solid line connects to the scale height of the youngest stellar particles at each epoch.
    The dashed blue and red lines are the scale heights ($h_{z}$) of the thin and thick disks derived from the double-component fit to the vertical profile measured at each epoch. 
    The vertical hatched band points to $z\sim1.7$, the time at which the disk structure begins to appear in this galaxy. 
    As the combined result of the thickening of the existing disk stars and the continued formation of young thin disk stars, the vertical distribution (and the scale heights of the thin and thick disks obtained as a result of the fit) does not change much since disk settling. This conspiracy points towards a confounding factor regulating simultaneously star formation and vertical diffusion.}
    \label{fig:hz_evolution}
\end{figure}

For a better visual guide, we measured the scale height of each mono-age population at each epoch using a single $\rm sech^{2}$ function and traced it with time in Figure~\ref{fig:hz_evolution}-(d).
Note that all the scale heights in this plot are measured in the cylindrical region at $2\,R_{\rm d}\pm1\,\rm kpc$ for the galaxy at each redshift to consider the size evolution of the galaxy. 
The evolution of the disk scale length ($R_{\rm d}$) is shown in Panel (c), and the scale length of this galaxy does not change much at least since $z\sim1.0$ \citep[c.f.,][]{Brook2006Universe}. 
We confirmed that the overall trend remains the same even when we measure all the scale heights at fixed cylindrical region (e.g., $R_{\rm xy}=3.4\pm1\,\rm kpc$), except for the first two earliest epochs ($z > 1.5$), at which the galaxy was much smaller.
We have adopted the same color key for ages from Figure~\ref{fig:vertical_profile_evolution}.
The scale heights of the thin and thick disks from the double $\rm sech^{2}$ fitting are shown as dashed blue and red lines. 
We applied the double-component fit, only after the galactic disk is developed in a galaxy. 
The disk settling epoch ($z_{\rm disk\,settling}$) is measured by the criterion using $V/\sigma$ of cold gas (mean $V/\sigma>3$ for $150\,\rm Myr$, see Section~\ref{sec: decomposition_kinematics} for details), and the evolution of $V/\sigma$ of cold gas is shown in Panel (b).
The epoch of the first emergence of the disk structure in the {\gal} galaxy is $z_{\rm disk\,settling}\sim1.7$ ($t_{\rm look\,back}\sim10\,\rm Gyr$), which is represented as the hatched vertical line. 
We also added the instantaneous star formation rate history of the galaxy in Panel (a).

The scale height of each mono-age population, represented as each colored solid line in Panel (d) of Figure~\ref{fig:hz_evolution}, increases with time, which is consistent with the kinematic heating of each mono-age population after birth \citep[e.g.,][]{Bird2020}.
The disk heating can be caused by several possible sources including minor mergers
\citep[e.g,][]{Quinn1993HeatingMergers, Kazantzidis2008ColdAccretion} and giant molecular clouds \citep[e.g,][]{Spitzer1951TheVelocities, Aumer2016Age-velocityGalaxies}. 
In addition, the distribution of each population not only becomes thicker with time but also gets fainter in $\sl r$-band with time (fading).
This is probably the reason why the slope of the increasing scale height seems to be the steepest right after each mono-age population was born; the distribution appears to be much thicker in luminosity than in mass due to fading.
Indeed, we confirm all the trends in this plot even when tracking the scale heights measured in {\em mass}, except that the thickening in the first two time steps becomes slightly smaller.

The youngest stellar particles at each epoch, i.e., the gray solid line in the bottom of Figure~\ref{fig:hz_evolution} (d), have a hint of gradually decreasing scale heights with time. 
The slight upturn in the gray line at $t_{\rm look\,back}\sim 3$-0 Gyr is due to the close encounter with two satellite galaxies. 
This encounter also enhances the star formation (Panel a) \citep[see also][]{Ruiz-Lara2020TheHistory} and at the same time acts as an external heating source. Note the drop of $V/\sigma$ in Panel (b) during the same period.

As the combined result of the thickening of the existing disk stars and the continued formation of young thin disk stars, the vertical profile of the disk itself does not change much over time. 
Therefore, the fitted scale heights of the thin and thick disks (blue and red dashed lines) remain almost constant since $z\sim1$ (lookback time $\rm \sim8\,Gyr$). 
This is consistent with the results of other studies \citep[e.g.,][]{Brook2006Universe}.
Furthermore, \cite{Elmegreen2006ObservationsField} found from the Hubble Ultra-Deep Field edge-on disk galaxies that their z850-band scale heights are nearly constant over a long period of time \citep[see also][]{Reshetnikov2019Edge-onField}. 
This conspiracy requires a high degree of synchronization between the efficiency of star formation on the one hand and the efficiency of vertical heating on the other.

Recall that star formation efficiency is partially driven by gas turbulence, which is closely impacted by the level of gravitational fluctuations within the disk. 
As shown in Dubois et al (in prep), the cosmic appearance of galactic disks in fact implies a fine-tuning between cooling and heating processes: cosmic evolution promotes a transition towards secularly-driven morphology which steers disks towards an effective Toomre $Q$ parameter (gas+star) close to one. 
In turn, this latter becomes an attractor value\footnote{In short, the closer $Q$ is to one, the shorter the feedback loop timescale, the tighter the loop.}.  
In the vicinity of $Q=1$, star formation is thus modulated by the dynamically induced level of turbulence in the gas disk. 
Indeed, several studies have shown that in a marginally stable disk, velocity dispersion of newly-formed young stars correlates with the surface star formation rate density \citep[e.g.,][]{Bird2020}.
In parallel, the efficiency of stellar orbital diffusion (combined radial migration and vertical heating) is also amplified by the square of the gravitational susceptibility \citep{Fouvry2017}, and is therefore strongly boosted near marginal stability. 
The net effect will be the joint recurrent formation of a new population of younger (dynamically cold) stars and the vertical diffusion and radial migration of the older population. 
This stratifies stars by age vertically, while preserving the double component vertical density profile of the existing disk, as is observed. Proximity to marginal stability is the confounding factor.

\subsection{The evolution of NH galaxies until $z\sim0.3$}
We have also investigated the evolution of the scale heights of the thin and thick disks and the mono-age groups of stellar particles in the NH disk galaxies down to $z\sim0.3$.
To explore the continuous evolution of the thin and thick disks over a considerable period of time, we selected 12 out of the 18 massive disk galaxies in the NH simulation which developed their disk structures at least by $z=1$ ($z_{\rm disk\,settling} > 1$) and maintained their disks until $z=0.3$ {\em without} experiencing violent events such as significant mergers (with fraction of ex-situ stars accreted after $z=1$ is lower than 0.1).
The purpose of this sub-sampling is to achieve reliable and stable measurements of disk properties. However, the overall measurements and conclusion stay the same even if we use all the 18 galaxies.
Since they have a quiescent mass growth history after $z=1$, their disk scale lengths have not changed much. 
The mean change in disk scale length between $z=1$ and $z=0.3$ is only by 18\%.

\begin{figure*}
    \linespread{1.0}\selectfont{}
    \includegraphics[width=\textwidth]{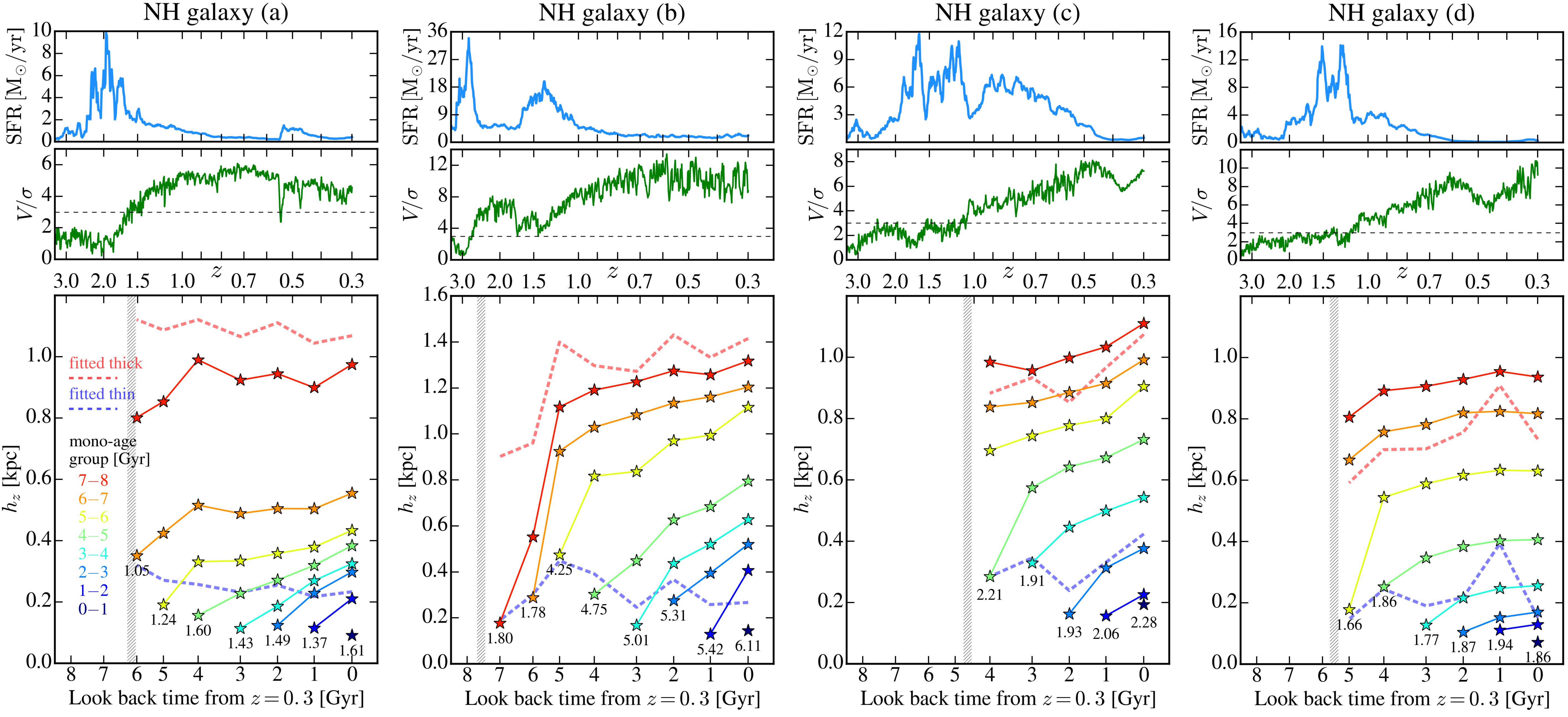}
    \caption{The same format as Figure~\ref{fig:hz_evolution}, but for the four NH galaxies down to $z=0.3$, as an example. The scale heights of the thin and thick disks are measured at $2\,R_{\rm d}$ of the galaxy at each epoch where $R_{\rm d}$ is the scale length of the disk which is displayed below each marker. The thickening of preexisting stars and the formation of young thin-disk stars contribute to the vertical distributions; as a result, the scale heights from the fit are broadly consistent since $z=1$ as long as galaxies keep forming young stars without having violent events such as mergers.}
    \label{fig:hz_evolution_NH}
\end{figure*}

We found that the scale heights of the thin and thick disks in these 12 galaxies evolve in response to the combined effect of disk thickening (distribution of existing older stars becoming thicker) and continued formation of disk stars in thin distributions. 
We brought the evolution of the four NH galaxies as examples in Figure~\ref{fig:hz_evolution_NH} in the same format as Figure~\ref{fig:hz_evolution}. 
We measured the vertical scale heights at $R=2\,R_{\rm d}\pm1\,\rm kpc$ at each epoch, and $R_{\rm d}$ at each epoch is listed below each marker. 
We trace the evolution of the thin and thick disks from the disk settling epoch measured using the same $V/\sigma$ (cold gas) criterion.

\begin{figure}
    \linespread{1.0}\selectfont{}
    \includegraphics[width=\columnwidth]{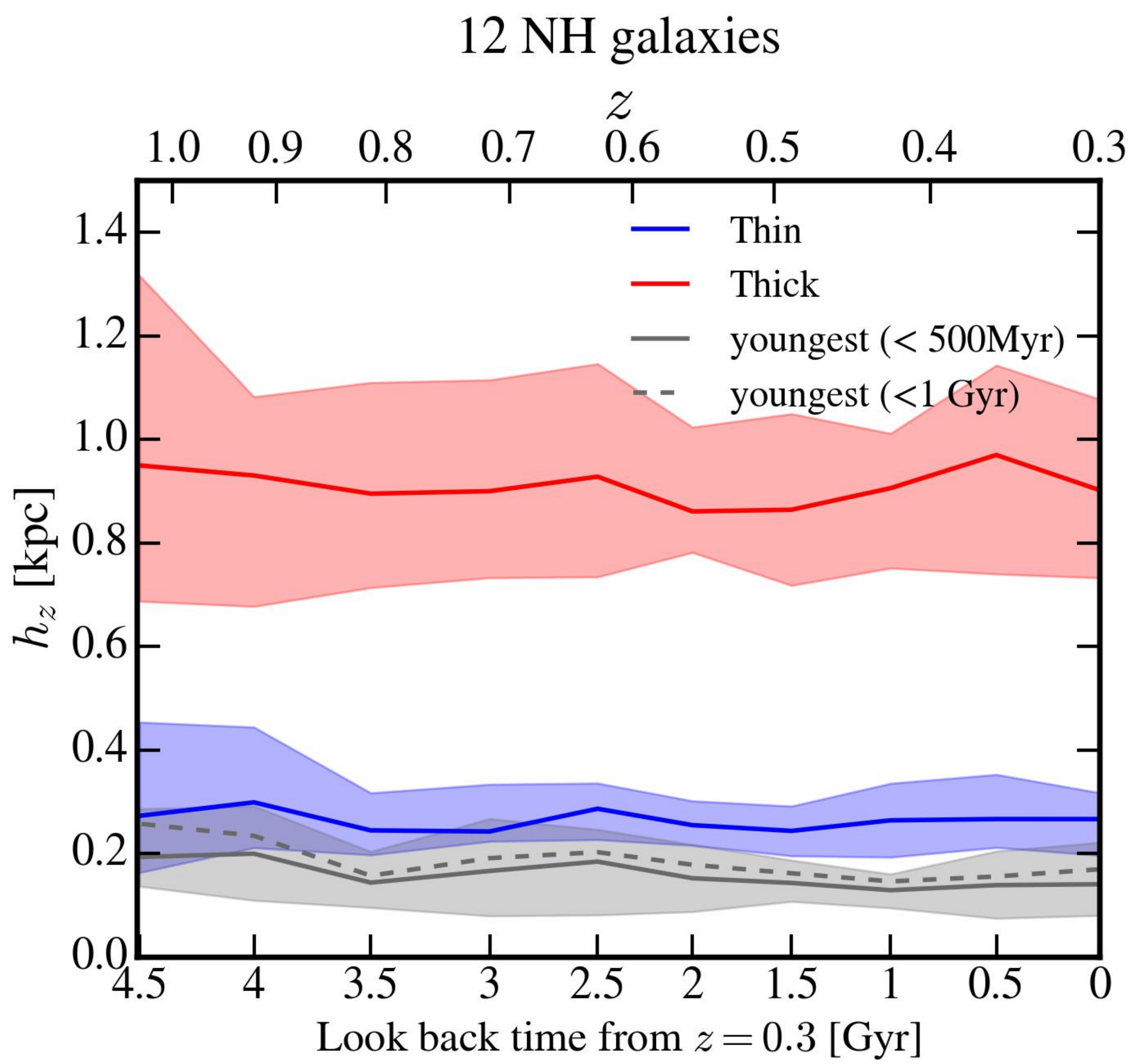}
    \caption{Time evolution of the median vertical scale heights of the thin (blue) and thick (red) disks in the 12 NH galaxies, from $z=1$ to $z=0.3$ (roughly 4--8 Gyr ago in look back time from $z=0.0$). The shades indicate the 16th to 84th percentiles of the vertical scale heights at each epoch. We selected 12 out of 18 disk galaxies in NH, which have settled their disks at least by $z=1$ and maintained their disks without experiencing significant mergers (with fraction of ex-situ stars accreted after $z=1$ is lower than 0.1).  
    The scale height of the youngest ($<0.5\,\rm Gyr$ or $<1.0\,\rm Gyr$) stars at each epoch is also displayed as a gray solid/dashed line for reference. We find no time evolution (at least since $z=1$) in the vertical distribution (and, thus, the scale heights from the fit) of the galaxies with a quiescent growth history.}
    \label{fig:10hz_evolution}
\end{figure}

For the NH galaxy (a), the galaxy started to build up its disk at $z\sim1.6$ (vertical hatched line), and since then, the scale heights of both thin and thick disks remain almost constant until $z\sim0.3$, just as the {\gal} galaxy presented in Figure~\ref{fig:hz_evolution}.

In the case of NH galaxy (b), its disk appeared in the galaxy since $z\sim2.6$ and the two-component fit to the vertical profile yields the scale heights of $\sim 250\,\rm pc$ and $\sim 850\,\rm pc$. 
Then it had two mergers at $z\sim1.5$ (with each mass ratio of $\sim$1:10), thus the star formation rate soared and the $V/\sigma$ of cold gas fell slightly at this epoch.
After the merger, scale heights of both thin and thick disks increased, partially due to the abrupt size growth of the galaxy; the scale length of the disk increased from $\rm 1.47\,kpc$ ($6\,\rm Gyr$ ago) to $4.25\,\rm kpc$ ($5\,\rm Gyr$ ago). 
Additionally, during this merger epoch, young stars were formed in a slightly thicker distribution out of more turbulent gas.
However, since $z\sim1$, there is no significant merger that could disrupt the disk or lead to any dramatic change in size or other galaxy properties.
Hence, the scale heights of both thin and thick disks do not change much until $z\sim0.3$. 
Also, the disk scale length increases by only 30\% during this period (from $4.75\,\rm kpc$ to $6.11\,\rm kpc$).

The same scenario applies to the NH galaxy (c). The scale heights are nearly constant between $z\sim1$ (after the galaxy developed its disk structure) to $z\sim0.5$. 
For the last $2\,\rm Gyr$, however, the scale heights of both disks increase as the galaxy does not form many young stars in that period; 
without the contribution of newly formed young stars to the total luminosity close to the midplane, the $\sl r$-band vertical distribution gets only thicker with time. 
The extreme case is the NH galaxy (d) where the galaxy is nearly quenched since $z\sim0.6$.
Between 1 to $\rm2\,Gyr$ ago, the overall vertical distribution became thicker and smoother.  
Only recently did the galaxy form a small number of new stars, and the thin disk with a sharp vertical distribution near the galactic midplane began to appear again, reducing both scale heights (see the two dashed lines).

Figure~\ref{fig:10hz_evolution} summarizes the scale height evolution of the 10 NH galaxies. The scale heights of the thin and thick disks virtually remain constant for the last $4\,\rm Gyr$ between $z=1.0$ and 0.3 (i.e., roughly $8\,\rm Gyr$ in look back time from the present epoch). 
This is consistent with the result from the {\gal} galaxy in the previous section. 
It seems that the vertical distribution of the disks in a galaxy and the scale heights of the two components from the fit do not change much since its disk settling epoch, provided that the galaxy does not experience serious external disturbances but maintains the star formation activity. 
Secular processes have taken over and regulate star formation and heating jointly, so as to maintain both scale heights.

It should be noted that numerical simulations are affected by the reliability of numerical computations. It is not entirely clear how numerical effects contribute to the disk heating \citep[e.g.,][]{House2011}. We can probably say that our simulations suffer less than many previous ones considering the substantially improved spatial and mass resolutions. Besides, most of the recent simulations seem to agree on the importance of heating in the disk evolution \citep[e.g.,][]{Aumer2016Age-velocityGalaxies, Grand2016VerticalContext}.
  
\section{Summary and Conclusion} 
\label{sec:conclusion}
The goal of this study was to see whether the spatially-defined thin and thick disks, derived from the vertical profile of a galaxy, are distinct components formed by different formation mechanisms. 
We used 18 massive disk galaxies (with $M_{\rm stellar}>10^{10}\,\msun$) from the {\nh} simulation that reached $z=0.3$ and one disk galaxy (with $M_{\rm stellar}=2.75\times10^{10}\,\msun$) from the {\gal} simulation that reached $z=0.0$. These simulations have unprecedentedly high spatial resolution (aimed to reach $\Delta x \sim 34\,\rm pc$ at $z=0.0$) for a relatively large volume simulation ({\nh}), which makes them ideal to investigate the detailed structures of galaxies.
The results of this study can be summarized as follows:

\begin{itemize}
\setlength\itemsep{0pt}
    \item We applied the widely accepted two-component fit to the $\sl r$-band vertical profiles, and the thin and thick disks were reasonably well reproduced in our {\gal} and NH simulations.
    The scale height/luminosity ratios between the thin and thick disks in the simulated galaxies, obtained from the fitting, are broadly consistent with observations. 
     
    \item We spatially decomposed the thin and thick disks and compared the properties of stellar particles in the two components (Table~\ref{table:properties}).
    Thick disks have a higher contribution of stellar particles formed ex situ, but both thin and thick disks are still dominated by in-situ formed stars.
    The thin disks in our simulated galaxies are younger, metal-richer, and rotating faster, which is qualitatively consistent with observations. 
    We found that different age distributions of the in-situ formed stars in the thin and thick disks contribute most to the different kinematics between the two components; while thin disks mostly consist of young stars with highly rotation-dominated kinematics, thick-disk stars are much older and rotating slower because some of them were formed with dispersion-dominated kinematics before disk settling, and others formed on the disks were also slowed down by heating.
    Roughly a half of the in-situ thick disk stars are formed on the disk after disk settling, but the exact number depends on detailed star formation and merger history of galaxies. 
    
    \item We traced the birthplace of stellar particles in each component and their kinematic properties at birth. 
    We found that a large part of the thick disk stars was spatially (formed close to the midplane) and kinematically (formed with higher orbital circularity) much thinner at birth.
    This suggests that the two disks are not entirely {\em distinct} in terms of formation process but they are rather snapshots in time and space of a {\em continuous} evolution of a galactic disk.
    
    \item Tracking the distribution of the mono-age populations, we found that the two-component disk (from the fit) is the result of the evolution of the disks where the galaxy continues to form young thin disk stars, while at the same time, the preexisting disks get thicker with time due to orbital diffusion. 
    As a result of the two effects, in most cases, the vertical distributions (scale heights) do not change much since $z\sim1$, as long as the galactic disks are in place and galaxies keep forming young stars without having violent events such as mergers. 
    The confounding factor tuning both the star formation of young (thin-disk) stars and the rate of vertical diffusion is the $Q\sim1$ attractor value (Dubois et al in prep), which accelerates both processes.
    
\end{itemize}

One important characteristic of thick disks that we have not directly covered in this study is their alpha abundances. 
It seems well established from many observations that MW disk stars show a bimodal distribution in the plane of [$\rm \alpha/Fe$]--[Fe/H], which are often associated with the thin and thick disks \citep[e.g.,][]{Lee2011FormationSample, Anders2014ChemodynamicsData, Bensby2014ExploringNeighbourhood, Duong2018TheNeighbourhood, Mackereth2019DynamicalitGaia}.
Also, there have been many studies using cosmological simulations of individual galaxies to explain the origin of this bimodality \citep[e.g.,][]{Brook2012ThinGalaxy, Grand2020SausageMerger, Buck2020OnMigration, Agertz2020VINTERGATANGalaxy, Renaud2020VINTERGATANMergers}. 
Since the simulations we used for this study did not trace the alpha elements separately, we instead investigated the age distribution of the stars in the thin and thick disk regions, and found that the two distributions appear to be different; for example, in Figure~\ref{fig:age_metal}, the age distribution of the thin-disk stars is skewed toward the younger ages (0--2 Gyr), while most of the thick disk stars were formed much earlier (8--12 Gyr ago).
So, it is important to note that, in terms of age (thus, formation epoch, and probably alpha abundances, as well), the two disks are distinct.

However, what we found in our study is that their \textit{formation processes} are not distinct; both disks are mostly formed in situ, and most of the stars in the thick disks are formed as thinner components when they were younger; they later evolved (migrated through heating) to become thicker components. 
In this regard, the spatially-separated two components cannot be directly compared with the chemically-distinct sequences in the plane of [$\rm \alpha/Fe$]--[Fe/H]. While chemically-defined sequences (whether stars are in an alpha-poor or alpha-enhanced sequence) are imprinted at birth (formation), spatially-defined components (whether they are thinner or thicker components) are the result of both formation {\it and} evolution.

We started this investigation by assuming that the break in the vertical mass or luminosity profiles of a disk galaxy indicates the presence of two separate components in the disk: i.e., thin and thick disks.  
Our simulations, however, show that spatially-defined thin and thick disks are {\em not entirely distinct} components in terms of {\em formation process}.
They are rather two parts of a single {\em continuous} disk component that evolves with time as a result of the continued star formation of thin-disk stars and disk heating.

\acknowledgments
We thank Cristina {Mart{\'\i}nez-Lombilla} for providing us with the data for Figure 4. We also thank the anonymous referee for useful comments that improved the clarity of the manuscript.
We thank Joss Bland-Hawthorn and Sanjib Sharma for useful information and discussion in the early stage of our paper writing.
S.K.Y., the corresponding author, acknowledges support from the Korean National Research Foundation (NRF-2020R1A2C3003769). The supercomputing time for numerical simulation was kindly provided by KISTI (KSC-2017-G2-003), and large data transfer was supported by KREONET, which is managed and operated by KISTI. 
This work is partially supported by grant Segal ANR-19-CE31-0017 of the French Agence Nationale de la Recherche: http://www.secular-evolution.org. We thank St{\'{e}}phane Rouberol for the smooth running of the GPUs on the Horizon cluster, where the simulations were performed. 
This work was granted access to the HPC resources of CINES under the allocations c2016047637, A0020407637 and A0070402192 by Genci, KSC-2017-G2-0003 by KISTI, and as a “Grand Challenge” project granted by GENCI on the AMD Rome extension of the Joliot Curie supercomputer at TGCC.
TK was supported in part by the National Research Foundation of Korea (NRF-2017R1A5A1070354 and NRF-2020R1C1C100707911) and in part by the Yonsei University Future-leading Research Initiative (RMS2-2019-22-0216).
The science of JD is supported by Adrian Beecroft and the STFC.

\bibliography{references}{}
\bibliographystyle{aasjournal}


\appendix
\section{Appendix}
\renewcommand{\thefigure}{A\arabic{figure}}
\setcounter{figure}{0}
\renewcommand{\thetable}{A\arabic{table}}
\setcounter{table}{0}

\begin{table*}[b]
\caption{Properties of the {\gal} ($z=0.0$) and {\nh} ($z=0.3$) galaxies.}
\begin{threeparttable}
\label{table:scale_height_and_luminosity_ratios}
\begin{adjustwidth}{-1.7cm}{}
       \makebox[1 \textwidth][c]{   
       \resizebox{1.12 \linewidth}{!}{

\begin{tabular}{p{0.1\textwidth}
>{\centering\arraybackslash}p{0.09\textwidth}
>{\centering\arraybackslash}p{0.06\textwidth}
>{\centering\arraybackslash}p{0.06\textwidth}
>{\centering\arraybackslash}p{0.06\textwidth}
>{\centering\arraybackslash}p{0.06\textwidth}
>{\centering\arraybackslash}p{0.1\textwidth}
>{\centering\arraybackslash}p{0.1\textwidth}
>{\centering\arraybackslash}p{0.1\textwidth}
>{\centering\arraybackslash}p{0.1\textwidth}
}
\hline
  Galaxy ID \tnote{(1)}
  & $M_{\rm stellar}$  \tnote{(2)}
  & $R_{\rm d}$  \tnote{(3)}
  & $R_{\rm 90}$  \tnote{(4)}
  & $V_{\rm circ}$  \tnote{(5)}
  & $B/T$  \tnote{(6)}
  & $z_{\rm thick}/z_{\rm thin}$  \tnote{(7)} 
  & $L_{\rm thick}/L_{\rm thin}$  \tnote{(8)}
  & $z_{\rm disk\,settling}$ \tnote{(9)}
  & $f_{\rm contam}$ \tnote{(10)} \\
  
  & $\,\rm [\times10^{10}\,\msun]$
  & [kpc]
  & [kpc]
  & [km/s]
  &
  &
  &
  &
  & [\%] \\
  
\hline
\hline

{\gal} \tnote{\textdagger} 
& $2.8$ 
& $1.7$
& $8.0$
& $192$ 
& $0.24$
& $4.06$
& $0.15^{+0.07}_{-0.08}$ 
& 1.7
& $0.00$\\ \hline

NH 1 \tnote{(b)}
& $7.7$ 
& $6.1$
& $22$
& $252$ 
& $0.21$
& $5.30$
& $0.20^{+0.23}_{-0.10}$ 
& 2.6
& $0.00$\\ 

NH 2  
& $5.4$
& $0.8$
& $8.0$
& $246$ 
& $0.32$
& $3.97$
& $0.20^{+0.06}_{-0.11}$ 
& 1.9
& $0.00$\\ 

NH 3  
& $5.3$ 
& $4.0$
& $23$
& $204$ 
& $0.21$
& $4.45$
& $0.12^{+0.14}_{-0.04}$ 
& 2.1
& $0.00$\\ 

NH 4 \tnote{(c)} 
& $3.6$ 
& $2.3$
& $11$
& $181$ 
& $0.19$
& $2.54$
& $0.44^{+0.20}_{-0.11}$ 
& 1.1
& $0.00$\\ 

NH 5 
& $2.7$ 
& $2.3$
& $9.1$
& $185$ 
& $0.20$
& $2.72$
& $0.39^{+0.11}_{-0.15}$ 
& 1.3
& $0.01$\\ 

NH 6  
& $2.6$ 
& $2.6$
& $12$
& $181$ 
& $0.21$
& $5.49$
& $0.39^{+0.29}_{-0.23}$
& 2.4
& $0.00$\\ 

NH 7 \tnote{(d)} 
& $2.4$ 
& $1.9$
& $7.9$
& $174$ 
& $0.21$
& $4.85$
& $1.48^{+0.44}_{-0.97}$ 
& 1.3
& $0.06$\\ 

NH 8  
& $2.2$ 
& $1.8$
& $12$
& $169$ 
& $0.29$
& $3.58$
& $0.08^{+0.07}_{-0.02}$ 
& 1.1
& $0.00$\\ 

NH 9 
& $1.8$ 
& $3.4$
& $17$
& $137$ 
& $0.14$
& $3.80$
& $0.70^{+0.40}_{-0.32}$ 
& 1.2
& $0.00$\\ 

NH 10  
& $1.5$ 
& $1.5$
& $8.7$
& $139$ 
& $0.22$
& $3.46$
& $0.13^{+0.04}_{-0.06}$ 
& 0.9
& $0.07$\\ 

NH 11 \tnote{(a)} 
& $1.5$ 
& $1.6$
& $6.5$
& $168$ 
& $0.39$
& $4.58$
& $0.32^{+0.22}_{-0.16}$ 
& 1.6
& $0.00$\\ 

NH 12  
& $1.3$ 
& $1.5$
& $7.5$
& $140$ 
& $0.29$
& $2.69$
& $0.17^{+0.06}_{-0.06}$ 
& 1.1
& $0.01$\\ 

NH 13  
& $1.3$
& $3.0$
& $11$
& $139$ 
& $0.15$
& $2.91$
& $0.34^{+0.14}_{-0.13}$ 
& 2.5
& $0.00$\\ 

NH 14  
& $1.2$ 
& $3.0$
& $13$
& $119$ 
& $0.22$
& $3.45$
& $0.69^{+0.53}_{-0.23}$ 
& 0.8
& $0.00$\\ 

NH 15  
& $1.1$ 
& $2.0$
& $9.2$
& $143$ 
& $0.20$
& $3.45$
& $0.50^{+0.14}_{-0.26}$ 
& 1.2
& $0.06$\\ 

NH 16 
& $1.1$ 
& $3.5$
& $13$
& $116$ 
& $0.20$
& $5.41$
& $0.43^{+0.27}_{-0.26}$ 
& 0.6
& $0.00$\\ 

NH 17  
& $1.1$ 
& $1.3$
& $10$
& $130$ 
& $0.32$
& $3.27$
& $0.55^{+0.12}_{-0.28}$ 
& 1.0
& $0.00$\\ 

NH 18 
& $1.0$ 
& $2.1$
& $9.6$
& $133$ 
& $0.22$
& $2.66$
& $0.30^{+0.06}_{-0.12}$ 
& 1.1
& $0.00$\\ \hline
\end{tabular}}}

\begin{tablenotes}[flushleft]
  \small
  \item[(1)] Galaxy IDs. The NH galaxies are listed in descending order of stellar mass.
  \item[(2)] $M_{\rm stellar}$: Stellar mass of galaxies. 
  \item[(3)] $R_{\rm d}$: Disk scale length of galaxies.
  \item[(4)] $R_{90}$: Radius containing 90\% of total stellar mass. 
  \item[(5)] $V_{\rm circ}$: Circular velocity of galaxies measured at $R_{90}$.
  \item[(6)] $B/T$: Bulge-to-total ratios where the ``bulge'' mass is defined as twice the mass sum of the stellar particles with $-0.5<\epsilon<0$. 
  \item[(7)] $z_{\rm thick}$/$z_{\rm thin}$: Scale height ratios between the thin and thick disks. 
  \item[(8)] $L_{\rm thick}$/$L_{\rm thin}$: Luminosity ratio between the two disks in $\rm |Z|<1\,kpc$ with errors showing from the ratios measured at \\
  the galactic midplane ($\rm |Z|=0\,kpc$) and the ratios measured in the regions ${\rm |Z|<2}\,z_{\rm thick}$.
  \item[(9)] $z_{\rm disk\,settling}$: Disk settling epoch defined as the time when the mean $V/\sigma$ of the cold gas inside $R_{90}$ of a galaxy for the past \\ $150\,\rm Myr$ reaches higher than 3.
  \item[(10)] $f_{\rm contam}$: Number fraction of low-resolution DM particles inside the halos in percentages.
  \item[\textdagger] {\gal} is investigated at $z=0.0$.
  \item[(a)] NH galaxy (a) in Figure~\ref{fig:hz_evolution_NH}
  \item[(b)] NH galaxy (b) in Figure~\ref{fig:hz_evolution_NH}
  \item[(c)] NH galaxy (c) in Figure~\ref{fig:hz_evolution_NH}
  \item[(d)] NH galaxy (d) in Figure~\ref{fig:hz_evolution_NH}
\end{tablenotes}
\end{adjustwidth}
\end{threeparttable}
\end{table*}

\end{document}